\documentclass[11pt,a4paper]{article}
\usepackage[hyperref]{eacl2021}
\usepackage{times}
\usepackage{latexsym}

\usepackage{microtype}

\usepackage{latexsym}
\usepackage{adjustbox}
\usepackage{appendix}
\usepackage{arydshln}
\usepackage{amsmath,amssymb}
\usepackage{array,booktabs,makecell}
\usepackage{tabularx}
\usepackage{adjustbox}
\usepackage{float}

\usepackage{multirow}
\usepackage[utf8]{inputenc}

\usepackage{multirow}
\usepackage{microtype}
\usepackage{xspace}
\usepackage{mathtools}
\usepackage{amssymb}
\usepackage{footnote}
\usepackage{tablefootnote}
\usepackage{graphicx}
\usepackage{color, colortbl}
\usepackage{xstring}
\usepackage{comment}
\usepackage{adjustbox}
\usepackage{float}
\restylefloat{table}
\usepackage{array,booktabs,makecell}

\usepackage{microtype}
\usepackage{graphicx}
\usepackage{subcaption}
\aclfinalcopy % Uncomment this line for the final submission

\usepackage{lipsum}
\usepackage{titlesec}

\title{Mega-COV: A Billion-Scale Dataset of 100+ Languages for COVID-19}

\author{Muhammad Abdul-Mageed, AbdelRahim Elmadany, El Moatez Billah Nagoudi, \\ \textbf{ Dinesh Pabbi, Kunal Verma, Rannie Lin} \\ 
   Natural Language Processing Lab\\ University of British Columbia \\
 \texttt{\small \{muhammad.mageed,a.elmadany,moatez.nagoudi\}@ubc.ca},\\ \texttt{  \{\small dinesh09,vkunal96\}@ece.ubc.ca},  \texttt{\small krieyalam@gmail.com}}

\date{}

\begin{document}
% \selectlanguage{english}
% \setcode{utf8}
\maketitle

\begin{abstract}
    We describe \texttt{Mega-COV}, a billion-scale dataset from Twitter for studying COVID-19. The dataset is diverse (covers 268 countries), longitudinal (goes as back as 2007), multilingual (comes in $100+$ languages), and has a significant number of location-tagged tweets ($\sim 169$M tweets). We release tweet IDs from the dataset. We also develop two powerful models, one for identifying whether or not a tweet is related to the pandemic  (best $F_1$=$97\%$) and another for detecting misinformation about COVID-19 (best $F_1$=$92\%$). A human annotation study reveals the utility of our models on a subset of \texttt{Mega-COV}. Our data and models can be useful for studying a wide host of phenomena related to the pandemic. \texttt{Mega-COV} and our models are publicly available. 
\end{abstract}
% \vspace*{-3mm}

% \vspace*{-5mm}
\setlength{\textfloatsep}{0.1cm}
\section{Introduction}\label{sec:intro}
%--------------------------------------
\noindent The seeds of the  coronavirus disease 2019 (COVID-19) pandemic are reported to have started as a local outbreak in Wuhan (Hubei, China) in December, 2019, but soon spread around the world~\cite{world2020statement}. As of January 24, 2021, the number of confirmed cases around the world exceeded $99.14$M and the number of confirmed deaths exceeded $2.13$M.\footnote{Source: The Center for Systems Science and Engineering, Johns Hopkins University. Dashboard: \url{https://www.arcgis.com/apps/opsdashboard/index.html\#/bda7594740fd40299423467b48e9ecf6}.} % The CSSE source provides real time updates on the location and number of confirmed COVID-19 cases.}  %The number of confirmed cases witnessed exponential growth in some countries. Overall, the growth has also been fast. For example, in April 7, when we first documented that number in the current manuscript, the number of cases was at 1,390,511.} 
In response to this ongoing public health emergency, researchers are mobilizing to track the pandemic and study its impact on all types of life in the planet. Clearly, the different ways the pandemic has its footprint on human life is a question that will be studied for years to come.
\begin{figure}[h]
  \centering
  \frame{\includegraphics[width=\linewidth]{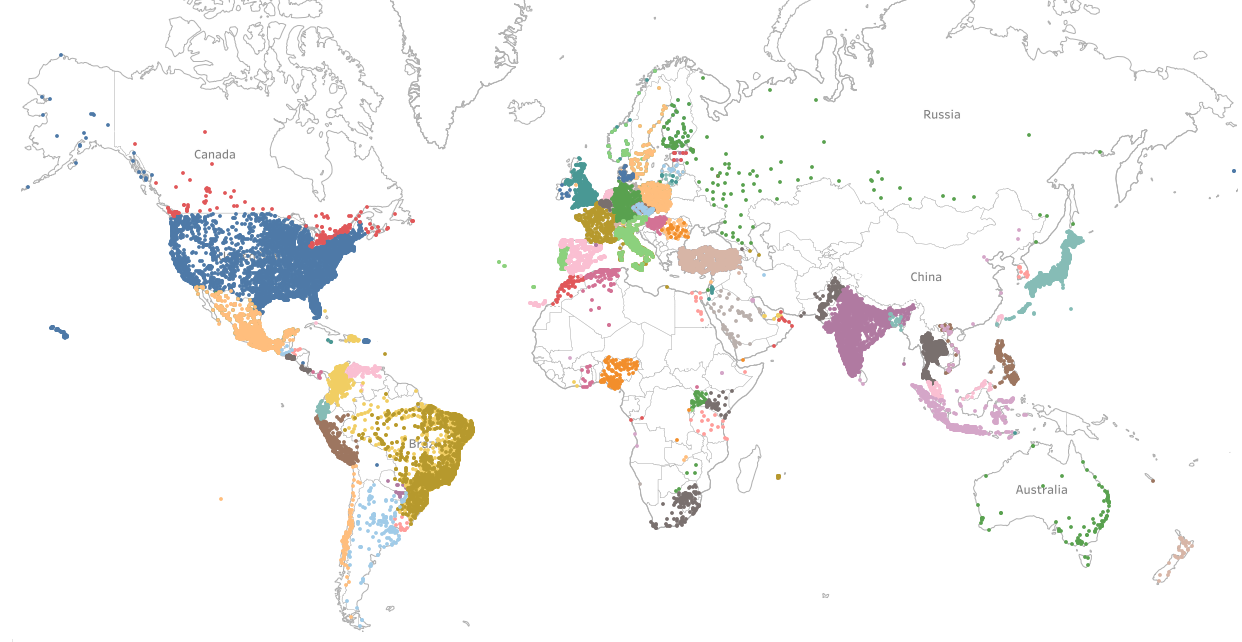}}
\caption{Global coverage of \texttt{Mega-COV} based on our geo-located data. Each dot is a city. Contiguous cities of the same color belong to the same country.}
\label{fig:geo-cities}
\end{figure}
Enabling scholarship on the topic by providing relevant data is an important endeavor. Toward this goal, we collect and release \texttt{Mega-Cov}, a billion-scale multilingual Twitter dataset with geo-location information.

As a result of the pandemic, most countries around the world went into lockdown and the public health emergency has restricted physical aspects of human communication considerably. As hundreds of millions of people spend more time sheltering in place, communication over social media became more important than ever. In particular, the content of social media communication promises to capture significant details about the lives of tens of millions of people. \texttt{Mega-Cov} is intended as a repository of such a content. %In this version of our work, the largest part of the dataset is focused on North America. Our next release will, however, bring a significant update with the size of the dataset doubling based on additional content from outside North America. 

There are several ongoing efforts to collect Twitter data, and our goal is to complement these. More specifically, we designed our methods to harvest a dataset that is unique in multiple ways, as follows:\\
\textbf{Massive Scale}: Very large datasets lend themselves to analyses that are not possible with smaller data. Given the global nature of COVID-19, we realize that a large-scale dataset will be most useful as the scale allows for slicing and dicing the data across different times, communities, languages, and regions that are not possible otherwise. For this reason, we dedicated significant resources to harvesting and preparing the dataset. \texttt{Mega-COV} has solid international coverage and brings data from $~$1M users from $268$ countries (see Section~\ref{subec:geo_div}). Overall, our dataset has $\sim$ 1.5B tweets (Section~\ref{sec:data}). This is one order of magnitude larger than \#COVID-19~\cite{chen2020covid}, the largest dataset we know of ($\sim144$M tweets as of June 1, 2020).\footnote{Both our own dataset and that of ~\newcite{chen2020covid} are growing over time. All our statistics in the current paper are based on our collection as of May 15, 2020. As of October 6, 2020, authors of \#COVID-19 report 649.9M tweets on their GitHub (\url{https://github.com/echen102/COVID-19-TweetIDs}), and \textit{our own dataset has exceeded 5B tweets}.}\\
\textbf{Topic Diversity}: We do not restrict our collection to tweets carrying certain hashtags. This makes the data general enough to involve content and topics directly related to COVID-19, regardless of existence of accompanying hashtags. This also allows for investigating themes that may not be directly linked to the pandemic but where the pandemic may have some bearings which should be taken into account when investigating such themes. This is important because users can, and indeed do, post about activities impacted by the health crisis without using any hashtags. In fact, users may not mention COVID-19 at all, even though what they are posting about could be affected by the pandemic one way or another (e.g., ``eating habits", ``shopping behavior"). Section~\ref{sec:append:hashtags} and Section~\ref{sec:append:domains} in the Appendix provide a general overview of issues discussed in the dataset.\\
\textbf{Longitudinal Coverage}: We collect multiple data points (up to 3,200) from each user, with a goal to allow for comparisons between the present \textit{and} the past across the same users, communities, and geographical regions (Section~\ref{subsec:temp-cov}). Again, this is desirable since without data from pre-COVID-19 time it will be challenging to hold any such comparisons. For example, some users may have stopped posting about ``exercising" during the pandemic but we cannot definitely identify this without access to these users' previous data where they may have been posting about their physical activities.\\
\textbf{Language Diversity}:
    Since our collection method targets users, rather than hashtag-based content, \texttt{Mega-COV} is \textit{linguistically diverse}. In theory, any language posted to Twitter by a user whose data we have collected should be represented. Based on Twitter-assigned language codes, we identify a total of $65$ languages. However, applying two different language detection tools to the whole dataset, we identify more than 100 languages.  (Section~\ref{subsec:ling-div}).\\
\textbf{No Distribution Shift}: Related to the two previous points, but from a machine learning perspective, by collecting the data without conditioning on existence of specific (or any) hashtags we avoid introducing distribution bias. In other words, the data can be used to study various phenomena in-the-wild. This warrants more \textit{generalizable} findings and models.\\
A dataset as large as \texttt{Mega-COV} can be hard to navigate. In particular, an informative description of the dataset is necessary for navigating it. In this paper, we provide an explanation of a number of global aspects of the dataset, including its geographic, temporal, and linguistic coverage. We also provide a high-level content analysis of the data, and explore user sharing of content from particular web domains with a focus on news media. In the context of our investigation of \texttt{Mega-COV}, we make an array of important discoveries. For example, we strikingly discover that, perhaps for the first time in Twitter history, \textit{users address one another and retweet more than they post tweets}. We also find a noticeable rise in ranks for news sites (based on how frequent their URLs are shared) during 2020 as compared to 2019, with \textit{a shift toward global (rather than local) news media}. A third finding is how \textit{use of the Twitter platform surged in March}, perhaps making it the busiest time in the history of the network. \\
Furthermore, we develop two groups of effective neural models: \textbf{(1) COVID-relevance models} (for detecting whether a tweet is related to COVID-19 or not). \textbf{(2)} \textbf{COVID-misinformation models} (for detecting whether a text carries fake information or not). In addition to releasing our best models, we also apply them to a total of 30M tweets from \texttt{Mega-COV} and release our tags to accelerate further research on the topic.%~\footnote{We plan to run the models on \textit{all} \texttt{Mega-COV} data and release the tags.} 

The rest of the paper is organized as follows: In Section~\ref{sec:data}, we describe our data collection methods. Section~\ref{sec:mega-cov-glance} is where we investigate geographic, linguistic, and temporal dimensions of our data. We describe our models for detecting COVID-19 tweets and COVID-misinformation in Section~\ref{sec:model}. Section~\ref{sec:megacov:applications} is where we apply our relevance and misinformation models to a large sample of \texttt{Mega-COV}. Section \ref{sec:ethics} is about data release and ethics. We provide a literature review in Section~\ref{sec:rel}, and conclude in Section~\ref{sec:conc}.

% We provide a case study of human mobility in Section~\ref{sec:hum-mob}. 

% \vspace*{-3mm}
\section{Data Collection}\label{sec:data}
%------------
%%%%%%%

%----------------
\noindent To collect a sufficiently large dataset, we put crawlers using the Twitter streaming API\footnote{API link:~\url{https://github.com/tweepy/}} on Africa, Asia, Australia, Europe, North America, and South America starting in early January, 2020. This allows us to acquire a diverse set of tweets from which we can extract a random set of user IDs whose timelines (up to 3,200 tweets) we then iteratively crawl every two weeks. This gave us data from July $30^{th}, 2020$ backwards, depending on how prolific of a poster a user is (see Table~\ref{fig:megacov-temp} for a breakdown.). In this paper, we describe and analyze the version of \texttt{Mega-COV} collected up to May 15, 2020 and use the term \texttt{Mega-COV} to refer it. \texttt{Mega-COV} comprises a total of $1,023,972$ users who contribute $1,487,328,805$ tweets. For each tweet, we collect the whole \texttt{json} object. This gives us access to various types of information, such as user location and the language tag (including ``undefined") Twitter assigns to each tweet. We then use the data streaming and processing engine, Spark, to merge all user files and run our analyses. To capture a wide range of behaviors, we keep \textit{tweets}, \textit{retweets}, and \textit{responses} (i.e., direct user-to-user interactions) as independent categories. Table~\ref{tab:main_data} offers a breakdown of the distribution of the different types of posts in \texttt{Mega-COV}. Tweet IDs of the dataset are publicly available at our GitHub\footnote{Accessible at: \url{https://github.com/UBC-NLP/megacov}.} and can be downloaded for research. To the extent it is possible, we intend to provide semi-regular updates to the dataset repository.
\begin{table}[]
\centering
\begin{adjustbox}{width=\columnwidth}{
\begin{tabular}{lrrrr}
\hline
\multicolumn{1}{l}{\textbf{Data}} & \textbf{Tweets} & \textbf{Retweets} & \textbf{Replies} & \textbf{All} \\ \hline
\textbf{2007-2020}      & 612M       & 507M         & 369M        & 1.5B    \\ 
\textbf{2020}          & 122M        &174M          & 129M         & 425M    \\ %\hline
\textbf{Users}       & 1M           & 976K             & 994K            & 1M        \\ \hline
\end{tabular}}
\end{adjustbox}
\caption{Distribution of tweets, retweets, and replies in \texttt{Mega-COV} (numbers rounded).}
\label{tab:main_data}
\end{table}

% \vspace*{-3mm}
\section{Exploring Mega-COV}\label{sec:mega-cov-glance}
%%%%%%%%%%%%%%%%%%%%%%%%%%%%%%%%%%%%%%%%%%%%%
%-------------
\begin{table*}[]
\resizebox{\textwidth}{!}{%
\centering
\begin{tabular}{lrrrrrrrr}
\cline{2-9}
\multicolumn{1}{c}{} & \multirow{2}{*}{\textbf{Geolocated}} &  & \textbf{Tweeted From} &  & \multirow{2}{*}{\textbf{Geotagged}}&  & \textbf{Tweeted From} & \\
\cline{3-5} \cline{7-9} 
\multicolumn{1}{c}{}                  &                                              & \textbf{Canada} & \textbf{U.S.} & \textbf{Other} &                                             & \textbf{Canada} & \textbf{U.S.} & \textbf{Other} \\ \hline
\textbf{All-Time}           & 186,939,854 &	16,459,655	& 70,756,282	& 99,723,917 &	31,392,563 &	3,600,952 &	11,449,400 &	16,342,211 \\ 
\textbf{All-Users}        & 739,645	& 102,388 &	327,213 &	463,673 &	266,916 &	47,622 &	117,096 &	165,860          \\ \cdashline{1-9} 

\textbf{2020}       & 65,584,908  & 3,331,720  & 24,259,973 & 37,993,215 & 2,942,675  & 246,185   & 1,187,131  & 1,509,359  \\
\textbf{2020-Users} & 670,314     & 61,205     & 254,067    & 392,627    & 109,348    & 14,525    & 43,486     & 62,837     \\ \hline
\end{tabular}
}
\caption{\texttt{Mega-COV} geolocated and geotagged users and their tweets from \texttt{North America} vs. \texttt{Other} locations. See also Table~\ref{tab:continent-dist} for statistics from top countries by continent. \\}
\label{tab:geo_data}
\end{table*}
%%%%%%%%%%%%%%%%%%%%%%%%%%%%%%%%%
\begin{table}[ht]
\centering
\begin{adjustbox}{width=\columnwidth}
\begin{tabular}{llrrrr}
\hline
\multirow{2}{*}{\textbf{Continent}}  & \multicolumn{1}{c}{\multirow{2}{*}{\textbf{Country}}} & \multicolumn{2}{c}{\textbf{All}}                                               & \multicolumn{2}{c}{\textbf{2020}}                                              \\  \cline{3-6} 
           &                                   & \multicolumn{1}{c}{\textbf{Geo-Located}} & \multicolumn{1}{c}{\textbf{Users}} & \multicolumn{1}{c}{\textbf{Geo-Located}} & \multicolumn{1}{c}{\textbf{Users}} \\ \hline

\multirow{5}{*}{Africa}       & Nigeria                               & 1,876,879                                 & 16,220                              & 1,057,742                                 & 14,872                              \\ 
                               & South   Africa                        & 1,503,181                                 & 9,751                               & 692,367                                   & 6,373                               \\  
                               & Egypt                                 & 873,079                                   & 8,840                               & 452,738                                   & 5,900                               \\  
                               & Ghana                                 & 373,996                                   & 3,942                               & 202,470                                   & 3,089                               \\  
                               & Kenya                                 & 373,667                                   & 4,480                               & 172,796                                   & 3,026                               \\ \hline 
                               
\multirow{6}{*}{Asia}          & Japan                                 & 7,646,901                                 & 32,038                              & 2,752,890                                 & 23,773                              \\ 
                               & Indonesia                             & 4,540,286                                 & 22,893                              & 1,871,154                                 & 18,056                              \\ 
                               & Spain                                 & 4,327,475                                 & 43,236                              & 1,431,567                                 & 20,902                              \\ 
                               & Philippines                           & 4,078,410                                 & 15,477                              & 1,636,265                                 & 11,011                              \\
                               & India                                 & 3,107,917                                 & 33,931                              & 1,576,549                                 & 27,940                              \\  
                               & Saudi   Arabia                        & 2,158,584                                 & 18,402                              & 833,634                                   & 15,087                              \\ \hline

Australia                      & Australia                             & 1,179,205                                 & 12,090                              & 352,215                                   & 5,454                               \\ \hline 

\multirow{5}{*}{Europe}        & UK                      & 11,714,012                                & 70,787                              & 2,970,848                                 & 44,420                              \\ 
                               & Turkey                                & 5,067,118                                 & 32,589                              & 1,463,550                                 & 25,477                              \\  
                               & France                                & 2,030,523                                 & 36,017                              & 729,500                                   & 12,497                              \\  
                               & Italy                                 & 1,829,369                                 & 27,071                              & 527,648                                   & 8,308                               \\  
                               & Germany                               & 1,272,339                                 & 24,215                              & 385,306                                   & 7,412                               \\ \hline

\multirow{3}{*}{North America} & US                       & 69,515,949                                & 327,213                             & 23,578,430                                & 254,067                             \\ 
                               & Canada                                & 16,066,337                                & 102,388                             & 3,200,804                                 & 61,205                              \\ 
                               & Mexico                                & 3,665,791                                 & 36,190                              & 1,106,352                                 & 17,406                              \\ \hline
\multirow{5}{*}{South America} & Brazil                                & 15,879,664                                & 48,339                              & 8,060,537                                 & 41,277                              \\ 
                               & Argentina                             & 3,142,778                                 & 14,576                              & 1,298,381                                 & 10,901                              \\ 
                               & Colombia                              & 1,612,765                                 & 10,319                              & 629,426                                   & 6,884                               \\  
                               & Chile                                 & 1,003,459                                 & 6,212                               & 378,770                                   & 3,674                               \\  
                               & Ecuador                               & 447,250                                   & 3,435                               & 170,098                                   & 2,221                               \\
 \hline
\end{tabular}%
\end{adjustbox}
\caption{Distribution of data over top countries per continent in \texttt{Mega-COV} (\textit{all data} vs. \textit{2020}).}
\label{tab:continent-dist}
\end{table}

%%%%%%%%%%%%%%%%%%%%%%%%%%%%%%%%%
\subsection{Geographic Diversity}\label{subec:geo_div}
%%%%%%%%%%%%%%%%%%%%%%%%%%%%%%%%%

\begin{figure}[]
  \centering
  \frame{\includegraphics[width=\linewidth]{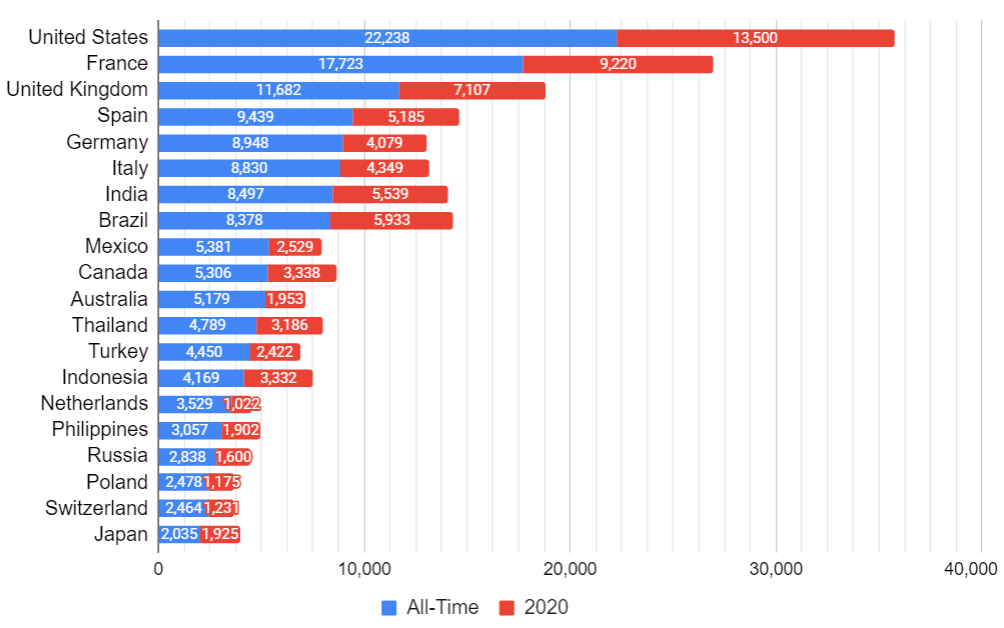}}
  \caption{Geographical diversity in \texttt{Mega-COV} based on \textit{geo-located} data.}
\label{fig:cities_2020_2019}%countries-with-topn-cities
\end{figure}

\noindent A region from which a tweet is posted can be associated with a specific `point' location or a Twitter place with a `bounding box’ that describes a larger area such as city, town, or country. We refer to tweets in this category as \textit{geo-located}. A smaller fraction of tweets are also \textit{geo-tagged} with longitude and latitude. As Table~\ref{tab:geo_data} shows, \texttt{Mega-COV} has $\sim187$M geo-located tweets from $\sim740$K users and $\sim31$M geo-tagged tweets from $\sim 267$K users. Table~\ref{tab:geo_data} also shows the distribution of tweets and users over the top two countries represented in the dataset, the U.S. and Canada (North America), and other locations (summed up as one category, but see also Table~\ref{tab:continent-dist} for countries in the data by continent). As explained, to allow comparisons over time (including behavioral changes during COVID-19), we include pre-2020 data in \texttt{Mega-COV}. For the year 2020, \texttt{Mega-COV} has $\sim 66$M geo-located tweets from $\sim 670$K users and $\sim3$M geo-tagged tweets from $\sim 109$K users.\footnote{The dataset has $\sim134$K  ``locations" which we could not resolve to a particular country using only the json information.} We note that significant parts from the data could still belong to the different countries but just not geo-located in the original \texttt{json} files retrieved from Twitter. Figure~\ref{fig:geo-cities}  shows actual point co-ordinates of locations from which the data were posted.  Figure~\ref{fig:cities_2020_2019} shows the geographical diversity in \texttt{Mega-COV} based on \textit{geo-located} data. We show the distribution in terms of the number of cities over \textit{the 20 countries from which we retrieved the highest number of locations} in the dataset, broken by \texttt{all-time} and the year \texttt{2020}. Overall,  \texttt{Mega-COV} has data posted from a total of $167,202$ cities that represent $268$ countries.  Figure~\ref{fig:geo-diversity} in Appendix~\ref{sec:append:data_geo} shows the distribution of data over countries. The top 5 countries in the data are \textit{the U.S., Canada, Brazil, the U.K.}, and \textit{Japan}. As we mention earlier, other top countries in the data across the various continents are shown in Table~\ref{tab:continent-dist}.
%%%%%%%%%%%%%%

\begin{figure}[]
  \centering
  \frame{\includegraphics[width=\linewidth]{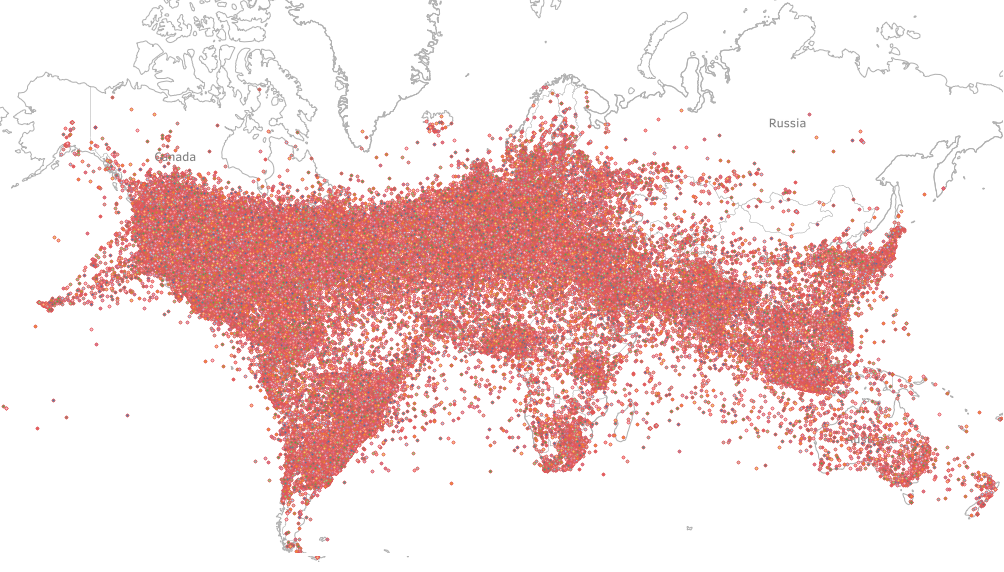}}
  %\caption{Geo-tagged coordinates}
\caption{World map coverage of \texttt{Mega-COV}. Each dot is a point co-ordinate (longitude and latitude) from which at least one tweet was posted. Clearly, users tweet while traveling, whether by air or sea.}
\label{fig:geo-cities}
\end{figure}

\begin{figure*}[th]
\begin{subfigure}{.485\linewidth}
  \centering
  \frame{\includegraphics[width=\linewidth]{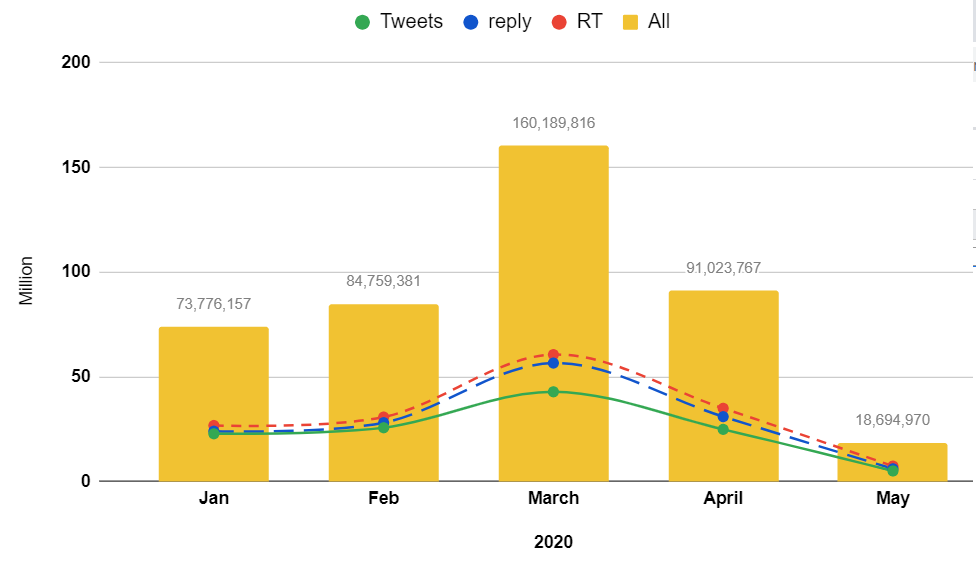}}
  \caption{Twitter user activity for Jan-May, 2020. }
  \label{fig:megacov-temp}
\end{subfigure}
\begin{subfigure}{.50\linewidth}
  \centering
  \frame{\includegraphics[width=\linewidth]{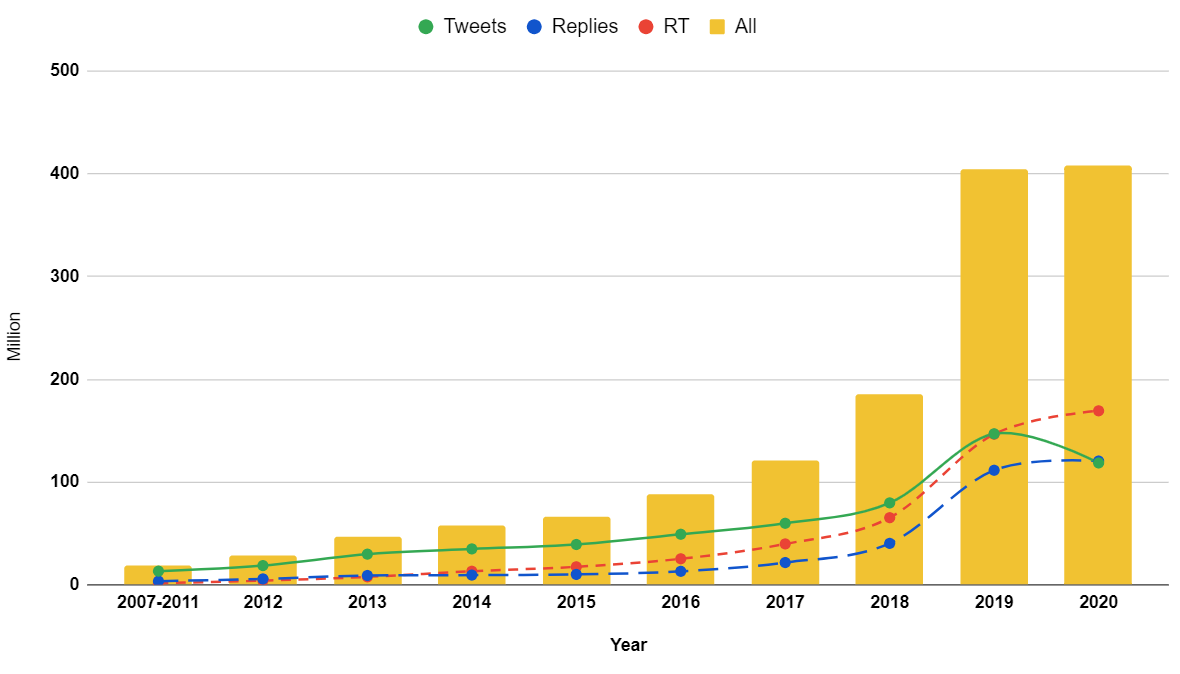}}
  \caption{Distribution of \texttt{Mega-COV} (2007-2020)}
  \label{fig:jan-may-posting}
\end{subfigure}
\caption{ Data distribution and user activity.}
\label{fig:inter-region}
\end{figure*}

%%%%%%%%%%%%%
\subsection{Temporal Coverage}\label{subsec:temp-cov}
%%%%%%%%%%%%%%%%%%%%%%%%%%%%%%
\noindent Our goal is to make it possible to exploit \texttt{Mega-COV} for comparing user social content over time. Since we crawl user timelines, the dataset comprises content going back as early as 2007. Figure~\ref{fig:megacov-temp} shows the distribution of data over the period 2007-2020. Simple frequency of user posting shows a surge in Twitter use in the period of Jan-April 2020 compared to the same months in 2019 %\footnote{The subset of \texttt{Mega-COV} we analyze here does not have enough data from the month of May to compare to.}
(see Figure~\ref{fig:avg-tweeting} in Appendix~\ref{sec:append:data_geo}). Indeed, we identify $40.53\%$ more posting during the first 4 months of 2020 compared to the same period in 2019. This is expected, both due to physical distancing and a wide range of human activity (e.g., ``work", ``shopping") moving online. More precisely, moving activities online causes users to be on their machines for longer times and hence have easier access to social media. The clear spike in the month of March 2020 is striking. It is particularly so given a \textit{shifted pattern of use: retweeting and replying (to others) are both observably more frequent than tweeting itself}. This especially takes place during the month of March, and somewhat continues in April, as shown in Figure~\ref{fig:jan-may-posting}. Figure~\ref{fig:megacov-temp} and Figure~\ref{fig:jan-may-posting} also show a breakdown of tweets, retweets, and replies. \textbf{\textit{A striking discovery is that, for 2020, users are engaged in conversations with one another more than tweeting directly to the platform.}} This may be the first time this pattern exists, perhaps in the history of the network. At least based on our massive dataset, this conclusion can be made. In addition, for 2020, we also see users retweeting more than tweeting. Based on \texttt{Mega-COV}, this is also happening for the first time. 
\subsection{Linguistic Diversity}\label{subsec:ling-div}
%%%%%%%%%%%%%%%%%%%%%%%%%%%%%%%%%%

\noindent We perform the language analysis based on tweets (n=$\sim 1.5$B), including retweets and replies.
 %~\footnote{This is an arbitrary decision, otherwise re-tweets and replies could be removed as relevant for some tasks.} 
Twitter assigns $65$ language ids to $\sim1.4$B tweets, while the rest are tagged as ``und" (for ``undefined"). \texttt{Mega-COV} has $\sim104$M ($\sim 7\%$) tweets tagged as ``und". We run two language identification tools, langid~\cite{lui2012langid} and Compact Language Detector~\cite{ooms2018cld2}\footnote{\url{https://code.google.com/p/cld2}.}  langid~\cite{lui2012langid},\footnote{\url{https://github.com/saffsd/langid.py}} on the whole dataset (including tweets tagged ``und" by Twitter).\footnote{As~\cite{lui2014accurate} point out, langid makes error on Twitter data. For this reason, we opted for adding predictions from CLD2.} %In addition, we are currently running equilid~\cite{jurgens2017incorporating} on the whole dataset and will release the labels from all the 3 tools for each tweet in \texttt{Mega-COV}.} 
After merging language tags from Twitter and the 2 tools, we acquire a total of $104$ labels. This makes \texttt{Mega-COV} very linguistically rich.  Table~\ref{tab:lang-dist-all} shows the top 20 languages identified by Twitter (left) and the top 20 languages tagged by one of the two tools, langid~\cite{lui2012langid}, \textit{after} removing the $65$ Twitter languages (right).

\begin{table}[ht]
\begin{adjustbox}{width=\columnwidth} 
\centering
\begin{tabular}{lr|lr}
\hline
\textbf{Lang} & \textbf{Freq} & \textbf{Lang} & \textbf{Freq} \\ \hline
English (en)	& 900M	& Hebrew (he)                & 1.2M\\
Spanish (es)	& 122M	& Croatian (hr)              & 685K\\
Portuguese (pt)	& 79.5M	& Maltese (mt)               & 325K\\
Japanese (ja)	& 46.6M	&Slovak (sk)                & 246K\\
Arabic (ar)	& 45M	& BI (id)      & 208K\\
Indonesion (in)	& 37M	& Latin (la)                 & 183K\\
French (fr)	& 29.5M	& Bosnian (bs)               & 143K\\
Turkish (tr)	& 28.5M	& Dzongkha (dz)              & 137.8K\\
Tagalog (tl)	& 19M	&Swahili (sw)               & 92K\\
Italian (it)	& 8.8M	& Azerbaijani (az)           & 68.9K\\
Thai (th)	& 7.7M	& Quechua (qu)               & 61K\\
Hindi (hi)	& 7M	& Albanian (sq)              & 61K\\
Dutch (nl)	& 6.9M	& Malay (ms)                 & 59K\\
Russuian (ru)	& 6.2M	& Kinyarwanda (rw)           & 56.8K\\
German (de)	& 6M	& Esperanto (eo)             & 55K\\
Catalan (ca)	& 3.5M	& Javanese (jv)              & 53K\\
Korean (ko)	& 2.9M	& Xhosa (xh) & 47.7K\\
Haitian Creole (ht)	& 2.8M	& Irish (ga) & 44.6K \\
Polish (pl)	& 2.4M	& Kurdish (ku) & 43K \\
Estonain (et)	& 2.1M	& Volapük (vo)  & 41K\\
\hline
\end{tabular}%
\end{adjustbox}
\caption{Top 20 languages assigned by Twitter (left) and top 20 languages assigned by langid (right) in \texttt{Mega-COV}. \textbf{BI:} Bahasa Indonesia.}
\label{tab:lang-dist-all}
\end{table}

\section{Models}\label{sec:model}
\noindent We develop two groups of models suited for answering important questions related to COVID-19, including making use of \texttt{Mega-COV}. These are \textbf{(1)~COVID-relevance}, where a classifier will label a tweet as \textit{relevant} to COVID-19 or \textit{not} and  \textbf{(2)~COVID-misinformation}, where a model predicts text veracity pertaining COVID-19 (i.e., whether a text carries \textit{true} or \textit{fake} information related to the pandemic). We now describe our methods.

\subsection{Methods}\label{subsec:methods}

\noindent For \textit{all} our models, we fine-tune 3 popular pre-trained language models: (1) Multilingual cased BERT (mBERT)~\cite{devlin2018bert} and (2-3) XLM Roberta base and large (XLM-R\textsubscript{Base}, XLM-R\textsubscript{Large})~\cite{conneau-etal-2020-unsupervised}.
The mBERT and XLM-R\textsubscript{Base} models have similar architectures, with $12$ layers each with $12$ attention heads, and $768$ hidden units. XLM-R\textsubscript{Large} has $24$ layers each with $16$ attention heads, and $1,024$ hidden units. While all the 3 models use a masking objective, the XLM-R models do not include the next sentence prediction objective used in BERT. 

\subsection{Hyper-Parameters and Optimization}\label{subsec:opt}
For each model, we use the same pre-processing in the respective code released by the authors. For \textit{all} models, we typically use a sequence length of $50$ tokens. We use a learning rate of $5$e$-6$ and a batch size of $32$. We train each model for $20$ epochs and identify the best epoch on a development set. We report performance on both development and test sets. We describe our baseline for each of the relevance and misinformation models in the respective sections below. We now introduce each of these two model groups.

\subsection{COVID-Relevance Models}

\noindent\textbf{Data.} Our COVID-relevance models predict whether a tweet is \textit{related} to COVID-19 or \textit{not} (i.e., not related). To train the models, we sample $\sim 2.3$M multilingual tweets (65 languages) collected with COVID-19 hashtags from~\cite{chen2020covid} and use them as our positive class (i.e., related to COVID-19). Examples of hashtags include \textit{\#Coronavirus}, \textit{\#covid-19}, and \textit{\#pandemic}. That is, we use the hashtags as a proxy for labels. This type of \textit{distant supervision} has been validated and widely used in many NLP models~\cite{go2009twitter,mohammad2015using,mageed2017emonet}. For the negative class (i.e., not related to COVID-19), we use a random sample of $\sim 2.3$M from the 2019 part (Jan-Nov) of \texttt{Mega-COV}. More description of the dataset we created for training the relevance models and the distribution of the data over the various languages is in Table~\ref{tab:append:relevance_data} (Appendix~\ref{sec:append:models}). 

\noindent\textbf{Splits and Training.} We split the data into 80\% TRAIN (n=3,146,334), 10\% DEV (n=393,567), and 10\% TEST (n=392,918). We then remove all hashtags which were used by \cite{chen2020covid} for collecting the data and fine-tune each of the 3 language models on TRAIN. %We use a sequence length of 50 tokens, the Adam optimizer with a learning rate of $5e^{-6}$, a batch size of 32, and run the models for 20 epochs. We identify the best epoch on the DEV set and report the respective model on both DEV and TEST. 

\noindent\textbf{Results.} As shown in Table~\ref{tab:res_covied_related}, XLM-R\textsubscript{Large} acquires best results with 97.95 \textit{acc} and 97.93 macro $F_1$ on TEST. These results are significantly better than a majority class baseline (based on TRAIN) and another arbitrarily-chosen (yet quite competitive) baseline model that chooses the related class (majority class in TRAIN) 75\% of the time.
% Train:  3,146,334 (3.2M)
% Dev: 393,567 (393.6K)
% test: 392,918 (393K)
% sequance length 50
% batch size 32
% lr 5e-06
% XML-Rlarge (3)
% XMLR base 6
% mBERT 5

% \usepackage{multirow}
% \usepackage{graphicx}

\begin{table}
\begin{adjustbox}{width=\columnwidth}

\renewcommand{\arraystretch}{1.3}
{\footnotesize
\begin{tabular}{lcccc}
\toprule
\multicolumn{1}{c}{}                              & \multicolumn{2}{c}{\textbf{DEV}} & \multicolumn{2}{c}{\textbf{TEST}} \\ \cline{2-5} 
\multicolumn{1}{c}{\multirow{-2}{*}{\textbf{Model}}} & \textbf{Acc}            & \textbf{ F$_{1}$}           & \textbf{Acc}           & \textbf{ F$_{1}$}            \\ \toprule
Baseline I  & $55.10$ & $71.05$ &  $54.99$& $70.96$ \\
Baseline II & $79.88$ & $88.81$ &  $75.33$& $85.93$ \\

\cdashline{2-5}
mBERT                                                                    & $97.35$                     & $97.33$                   & $97.39$                     & $97.37$                    \\ 
XLM-R\textsubscript{Base}                                                         & $97.72$                      & $97.70$                   & $97.71$                      & $97.69$                     \\ 
XLM-R\textsubscript{Large}                                                        & \textbf{$\bf97.92$}             & \textbf{$\bf97.90$}           & \textbf{$\bf97.95$}             & \textbf{$\bf97.93$}            \\ \toprule
\end{tabular}}
\end{adjustbox}
\caption{Performance of COVID-relevance models. \textbf{Baseline I:} Majority class in TRAIN. \textbf{Baseline II:} A model that chooses the majority class (related class) 75\% of the time.}
\label{tab:res_covied_related}

\end{table}

%%%%%%%%%%%%%%%

\noindent \textbf{Model Generalization.}
Our COVID-relevance models are trained with distant supervision (hashtags as surrogate labels). It is conceivable that content related to COVID-19 would still occur in real world without accompanying hashtags. To test the extent to which our best model would perform on external data, we evaluate it on two external Twitter datasets, CoAID~\cite{cui2020coaid} and ReCOVery~\cite{zhou2020recovery}, both of which are claimed by the authors to be completely (100\%) related to COVID-19.\footnote{Each of the two datasets are also labeled for fake news (\textit{true} vs. \textit{fake}) focused on COVID-19, but our focus here is exclusively on using the two datasets as gold-labeled TEST sets for evaluating our COVID-relevance model. Note that we will use these two datasets again as explained in Section~\ref{subsec:COVID-Misi} as well.}

As Tabel~\ref{tab:eval_COVID_related_on_Fake} shows, We do observe a drop in model performance as compared to our best model on our own TEST set in Table~\ref{tab:res_covied_related} (acc drops on average by $15.5$\% and $7.6\%$ $F_1$). However, the best model is still highly effective. It acquires an average acc of $82.46\%$ and $F_1$ of $90.38\%$ on the CoAID and ReCOVery datasets. We now introduce our misinformation models.

\begin{table}[ht]
\centering
\begin{adjustbox}{width=4.5cm}

\renewcommand{\arraystretch}{1.2}
{\footnotesize

\begin{tabular}{lcc}
     \toprule     
     
     \multirow{1}{*}{\textbf{Data}}    &    \textbf{Acc}    &  \textbf{F$_{1}$}        \\   \toprule 
\textbf{COAID}          &    $76.25$   &  $86.52$    \\ %\cline{2-5}
  \textbf{ReCOVery}  &                     $89.46$ &        $94.44$   \\ \hline
  
    \textbf{Average}  &                   $82.46$ &        $90.38$   \\
  \toprule
\end{tabular}}
\end{adjustbox}
\caption{Performance of our COVID-relevance models on the Twitter data in CoAID,  ReCOVery, and CoAID+ReCOVery. }
\label{tab:eval_COVID_related_on_Fake}
\end{table}

\subsection{COVID-Misinformation Models}
\label{subsec:COVID-Misi}
\noindent To train models for detecting the veracity of news related to COVID-19, we exploit  two recent and publicly available fake news  datasets (in English):   CoAID~\citep{cui2020coaid}, and ReCOVery~\citep{zhou2020recovery}. We now describe each of these datasets:

\begin{table}[ht]
\begin{adjustbox}{width=\columnwidth}

\renewcommand{\arraystretch}{1.4}
{\footnotesize

\begin{tabular}{lcccccc}
\toprule 
          & \multicolumn{3}{c}{\textbf{Fake}}                                                       & \multicolumn{3}{c}{\textbf{True}}                                                        \\   \cline{2-7}
          &  \textbf{Claims}                   & \textbf{News}                     & \textbf{Tweets}                     & \textbf{Claims}                   & \textbf{News}                     & \textbf{Tweets}                      \\   \toprule
\textbf{CoAID}     & 839             & 837  & 10,900        & 376    & 2716 & 149,343            \\ 
\textbf{ReCOVery}  &    -  & 665        &               26,418              &        -   & 1,364     & 114,402    \\  \toprule                
%\textbf{FakeCovid} &       -          & 6,254         &        -                   &    -                     & 1,369                   &         -                   \\ 
\textbf{Total}     & 839 & 1,502 & 37,318 & 376 & 4,080 & 263,745 \\ \toprule

\end{tabular}}
\end{adjustbox}
\caption{COVID-19 Misinformation Datasets.}\label{tab:fake_data}
% \vspace{-4mm}
\end{table}

\begin{table}[ht]
\begin{adjustbox}{width=\columnwidth}

\renewcommand{\arraystretch}{1.3}
{\footnotesize
\begin{tabular}{lcccccc}
\toprule
                     & \multicolumn{6}{c}{\textbf{Tweets}}                                                           \\ \cline{2-7}
                     & \multicolumn{3}{c}{\textbf{Fake}}                      & \multicolumn{3}{c}{\textbf{True}}                      \\\toprule
                     & \textbf{TRAIN} & \textbf{DEV} & \textbf{TEST} & \textbf{TRAIN} & \textbf{DEV} & \textbf{TEST} \\\toprule
\textbf{CoAID}       & 8,072          & 1,009        & 1,009         & 110,076        & 13,759       & 13,759        \\
\textbf{ReCOVery}    & 18,272         & 2,284        & 2,284         & 86,437         & 10,805       & 10,805        \\\cline{2-7}
\textbf{CoAID*}       & 8,072          & 163        & 171        & 110,076        & 6,314       &  6,388        \\ 
\textbf{ReCOVery*}    & 18,272         & 154        & 139        & 86,437         & 1,218      & 1,263        \\
%\textbf{FakeCovid}   & -              & -            & -             & -              & -            & -         \\
\toprule    
\end{tabular}}
\end{adjustbox}
\caption{\small {Statistics of CoAID and ReCOVery datasets across the data splits. CoAID$^{*}$ and  ReCOVery$^{*}$  are de-duplicated versions.}  }\label{tab:data_splits_tw}
\end{table}

\noindent \textbf{CoAID.} \citet{cui2020coaid} present a \textbf{C}ovid-19 he\textbf{A}thcare m\textbf{I}sinformation \textbf{D}ataset
(CoAID), with diverse COVID-19 healthcare misinformation, including fake news on websites and social platforms, along with related user engagements (i.e., tweets and replies) about such news. CoAID includes $3,235$ news articles and claims, $294,692$ user engagement, and $851$ social platform posts about COVID-19.  The dataset is collected from December 1, 2019 to July 1, 2020. Table~\ref{tab:fake_data} shows class distribution of news articles and tweets in CoAID.  More information about CoAID is in Appendix~\ref{sec:append:misinfo}.% in the Appendix.

%The topics of  CoAID include: \textit{\{COVID-19, coronavirus, pneumonia, flu9, lock down,stay home, quarantine and ventilator\}}.

\noindent \textbf{ReCOVery.} 
 \newcite{zhou2020recovery} choose $60$ news publishers with `extreme' levels of credibility (i.e., \textit{true} vs. \textit{fake} classes) from an original list of $\sim2,000$ to collect a total of $2,029$ news articles on COVID-19, published between January and May $2020$. They also collect $140,820$ tweets related to the news articles, considering those tweets related to true articles to be true and vice versa. Table~\ref{tab:fake_data} shows class distribution of news articles and tweets in ReCOVery.
 
\noindent \textbf{Splits and Cleaning.} 
Table~\ref{tab:data_splits_tw} shows the distribution of tweets in CoAID and Recovery \textit{before} and \textit{after} the de-duplication process. As  Table~\ref{tab:data_splits_tw} shows, de-duplication results in significantly reducing the sizes of DEV and TEST sets in the two resources. The  distribution of  news article  is shown in Table~\ref{tab:data_splits} (Appendix~\ref{sec:append:misinfo}).

\noindent\textbf{Training.} We  use  both  CoAID and ReCOVery after de-duplication for training neural models to detect fake news related to Covid-19.  Using the same hyper-parameters and training setup as the COVID-relevance models, we fine-tune the pre-trained language models on the Twitter dataset and the news dataset, independently.\footnote{Even though we could have used the monolingual versions of the transformer-based language models (i.e., BERT and RoBERTa), we stick to the multilingual versions for consistency.} Since \texttt{Mega-COV} is a social media dataset, we only focus on training Twitter models here and provide the news models in Appendix~\ref{sec:append:misinfo}. For the \textit{Twitter models}, we develop one model on CoAID, another on ReCOVery, independently, and a third model for  CoAID+ReCOVery (concatenated). Again, for each of these 3 datasets, we fine-tune on TRAIN and identify the best model on DEV. We then report the best model on both DEV and TEST. %For the current models, we use the majority class in TRAIN (\textit{true} class) as our baseline.\footnote{Note that this is a competitive baseline since the class distribution is very biased.}

\noindent\textbf{Results.} Since our focus is on detecting \textit{fake} texts, we show results on the positive class only in Table~\ref{tab:res-fake_only}. We report results in terms of \textit{precision}, \textit{recall} and \textit{F}$_1$. Our baseline is a small LSTM with 2 hidden layers, each of which has $50$ nodes. We add a dropout of $0.2$ after the first layer and arbitrarily train the LSMT for $3$ epochs. As Table~\ref{tab:res-fake_only} shows, our best results for fake tweet detection on TEST for CoAID is at $90\%$ $F_1$ (mBERT/XLM-R\textsubscript{Large}), for ReCOV $68\%$ (mBERT), and for these two combined is $92\%$. All results are above the LSTM baseline. We show results of the COVID-misinformation \textit{news} models in Table~\ref{tab:append:res-fake-news} (Appendix~\ref{sec:append:misinfo}).

\begin{table}[ht]
 	
\begin{adjustbox}{width=\columnwidth}
\renewcommand{\arraystretch}{1.5}{
\centering
\begin{tabular}{llcccccc}

\toprule
 \multirow{2}{*}{\textbf{Data}}  & \multirow{2}{*}{\textbf{  Model}} & \multicolumn{3}{c}{\textbf{DEV}}     & \multicolumn{3}{c}{\textbf{TEST}}       \\  \cline{3-8} 
      &    &   \textbf{Precision} & \textbf{Recall}& \textbf{ F$_{1}$}         & \textbf{Precision} & \textbf{Recall}&    \textbf{ F$_{1}$}        \\\toprule

   \multirow{4}{*}{ \textbf{CoAID}}    & LSTM     & $81.00$& $91.00$ & $86.00$ & $95.00$ & $78.00$  & $86.00$\\
   \cdashline{2-8}
        & mBERT     & $91.00$& $84.00$ & $87.00$ & $94.00$ & $87.00$  & $\bf90.00$\\

     &     XLM-R\textsubscript{Base}   & $93.00$& $\bf87.00$ & $90.00$ &$87.00$ & $88.00$  & $88.00$\\ 
  &      XLM-R\textsubscript{Large}         & $\bf98.00$& $86.00$ & $\bf92.00$ &$\bf97.00$ & $\bf93.00$  & $\bf90.00$\\
 % &    &      AraBERT        & $62.03$	&$47.72$	&$61.62$	&$49.27$ \\  \cline{2-6}
  
\toprule 
%%%%%%%%%%%%%%%%%%%%%%%%%%%%%%%%%%%%%%%%%%%%%%%%%%%%%%%%%%%%%%%%%  

      \multirow{4}{*}{ \textbf{ReCOV}}    & LSTM     & $60.00$& $56.00$ & $58.00$ & $54.00$ & $57.00$  & $55.00$\\
   \cdashline{2-8}
   & mBERT      & $81.00$& $\bf59.00$ & $\bf68.00$ &$87.00$ & $\bf55.00$  & $\bf68.00$\\
  &     XLM-R\textsubscript{Base}   &$72.00$& $58.00$ & $64.00$ &$75.00$ & $\bf55.00$  & $64.00$\\
      &      XLM-R\textsubscript{Large}      & $\bf89.00$& $52.00$ & $66.00$ &$\bf89.00$ & $51.00$  & $65.00$\\

%%%%%%%%%%%%%%%%%%%%%%%%%%%%%%%%%%%%%%%%%%%%%%%%%%%%%%%%%%%%%%%%%  

 \toprule

     \multirow{4}{*}{ \textbf{CoAID+ReCOV}}    & LSTM     & $79.00$& $58.00$ & $67.00$ & $66.00$ & $70.00$  & $68.00$\\
   \cdashline{2-8}
   & mBERT     & $\bf94.00$& $89.00$ & $\bf91.00$ &$\bf94.00$ & $89.00$  & $\bf92.00$\\ 
     &     XLM-R\textsubscript{Base}  & $88.00$& $88.00$ & $88.00$ &$88 .00$ & $88.00$  & $88.00$\\
     &      XLM-R\textsubscript{Large}     & $86.00$& $\bf94.00$ & $90.00$ &$85.00$ & $\bf93.00$  & $89.00$\\
 % &    &      AraBERT        & $62.03$	&$47.72$	&$61.62$	&$49.27$ \\  \cline{2-6}
  
\cline{3-6}

%%%%%%%%%%%%%%%%%%%%%%%%%%%%%%%%%%%%%%%%%%%%%%%%%%%%%%  

%%%%%%%%%%%%%%%%%%%%%%%%%%%%%%%%%%%%%%%%%%%%%%%%%%%%%%  

  \toprule

\end{tabular}
} \end{adjustbox}

\caption{\small  Performance of our COVID-misinformation \textit{Twitter} models  on the fake class only across the 3 settings CoAID, ReCOVery, and CoAID+ReCOVery. LSTM is our baseline.}
\label{tab:res-fake_only}
\end{table}

%%%%%%%%%%%%%%%%%%%%%%%%%%%
\section{Applications on Mega-COV}\label{sec:megacov:applications}
\noindent Now that we have developed two highly effective models, one for COVID-relevance and another for COVID-misinformation, we can employ these models to make discoveries using \texttt{Mega-COV}. Since our misinformation models are focused only on English (due to the external gold data we used for training being English only), we will restrict this analysis to the English language.\footnote{But we emphasize the multilingual capacity of our COVID-relevance model.} We were curious whether model predictions will have different distributions on the different types of Twitter posts (i.e., tweets, retweets, and replies). Hence, to enable such comparisons, we extract a random sample of $10$M samples from each of these post types (for a total of $30$M) from the year 2020 in \texttt{Mega-COV}. We then apply the XLM-R\textsubscript{Large} \textit{relevance} and \textit{misinformation} models on the extracted samples. Table~\ref{tab:eval_models_on_30M} shows the distribution of predicted labels from each of the two models across the 3 posting types (tweets, retweets, and replies). Strikingly, as the top half of the table shows, while only $7.77\%$ of tweets are predicted as \textit{COVID-related}, almost all retweets (99.84\%) are predicted as \textit{related}. This shows that users' retweets were focused almost exclusively on COVID-19. The table (bottom half) also shows that retweets are highest carriers of content predicted as \textit{fake} ($3.67\%$), followed by tweets ($2.3\%$). From the table, we can also deduce that only $2.45\%$ of all English language Twitter content (average across the 3 posting types) are predicted as fake. \textit{\textbf{Given the global use of English, and the large volume of English posts Twitter receives daily, this percentage of fake content is still problematically high.}}  

\subsection{Annotation Study}
\noindent We perform a human annotation study on a small sample of $150$ random posts from those the model predicted as both COVID-related \textit{and} fake. Two annotators labeled the $150$ samples for two types of tags, relevance and veracity. For relevance, all the $150$ posts were found relevant by the two annotators (perfect agreement). For veracity, since some posts can be very challenging to identify, we asked annotators to assign one of the $3$ tags in the set \textit{\{true, fake, unknown\}}. We did not ask annotators to consult any outside sources (e.g., Wikipedia or independent fact-checking sites) to identify veracity of the samples. Inter-annotator agreement is at \textit{Kappa (K)}=$77.81\%$, thus indicating almost perfect agreement. On average, annotators assigned the \textit{fake} class 39.39\% of the time, the \textit{true} class 3.02\%, and the \textit{unknown} class 57.05\%. While these findings show that it is hard for humans to identify data veracity without resorting to external sources, it also demonstrates the utility of the model in detecting actual fake stories in the wild. We provide a number of samples from the posts that were automatically tagged as \textit{COVID-related} and either \textit{true} or \textit{false} by our misinformation/veracity model in Table~\ref{tab:COVID-Misin-examples}. 
\begin{table}[ht]
\centering
\begin{adjustbox}{width=6.8cm}
\renewcommand{\arraystretch}{1.3}
{\footnotesize
\begin{tabular}{cllr}\toprule
\textbf{Model}   & \textbf{Data}  & \textbf{Prediction }&\textbf{Percentage}  \\ \toprule
          &   \multirow{2}{*}{{Tweets}} \textbf{}    &   Related & $7.77$  \\

  \multirow{3}{*}{\textbf{COVID}}          &           & Unrelated & $92.23$ \\ \cline{2-4} 
   \multirow{3}{*}{\textbf{Relevance}}           &    \multirow{2}{*}{{Retweets}}  & Related  &$99.84$ \\

            &           & Unrelated & $0.16$  \\ \cline{2-4} 
        &  \multirow{2}{*}{{Replies}}       & Related & $12.94$  \\

            &           & Unrelated  & $87.06$ \\ \toprule 

               & \multirow{2}{*}{{Tweets}}        & {Fake}&  $2.3$ \\
        \multirow{3}{*}{\textbf{COVID}}           &           & {True}& $97.10$  \\ 
        \cline{2-4} 
    \multirow{3}{*}{\textbf{Misinfo.}}           &  \multirow{2}{*}{{Retweets}} & {Fake}&  $1.38$\\
           &                & {True} &  $98.33$ \\  \cline{2-4} 
            &  \multirow{2}{*}{{Replies}}        & {Fake}&  $3.67$  \\
            &           & {True}&  $96.62$ \\\toprule
\end{tabular}} \end{adjustbox}
\caption{Distribution of predicted labels from our COVID-relevance  and COVID-misinformation models on randomly selected $30$M English samples from \texttt{Mega-COV} data. }
\label{tab:eval_models_on_30M}
\end{table}

\begin{table*}[ht]
 	
\small
\begin{adjustbox}{width=16cm}
\renewcommand{\arraystretch}{1.7}{
\small
\centering
\begin{tabular}{lc}
\toprule
\multicolumn{1}{c}{\bf Post}                                                                                                                                    & \textbf{Prediction}               \\
\toprule
Vatican confirms Pope Francis and two aides test positive for Coronavirus - MCM Whoaa 	\includegraphics[height=1.1em]{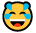} URL                                              & Fake                     \\
\includegraphics[height=1.1em]{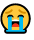}$\sim$CDC recommends men shave their beards to protect against coronavirus – USER URL                                                  & Fake                     \\
COVID - 19 : Chinese health authorities confirm patient zero ' had sex with bats ' URL                                                    & Fake                     \\
Royal Palace confirms Queen Elizabeth tests positive for coronavirus URL                                                                  & Fake                     \\ \hline
Is COVID - 19 airborne contagious ? New study shows that coronavirus may be caught from the air * 3 - hours * after it has been exposed . & \multicolumn{1}{c}{True} \\
A close relative of SARS-CoV - 2 found in bats offers more evidence it evolved naturally URL                                              & \multicolumn{1}{c}{True} \\
Antiviral remdesivir prevents disease progression in monkeys with COVID - 19 | National Institutes of Health ( NIH ) URL                  & \multicolumn{1}{c}{True} \\
COVID Surges Among Young Adults URL                                                                                                       & \multicolumn{1}{c}{True} \\
\toprule
\end{tabular}} \end{adjustbox}
\caption{\small Sample \texttt{Mega-COV} posts predicted as \textit{COVID-related}, and either \textit{true} or \textit{fake} by our models. }
\label{tab:COVID-Misin-examples}
\end{table*}

% \vspace*{-3mm}
\section{Data Release and Ethics}\label{sec:ethics}
%%%%%%%%%%%%%%%%%%%%%%%%%%%%%%%%%%%%%%
\noindent\textbf{Data Distribution.} The size of the data makes it an attractive object of study. Collection and exploration of the data required significant computing infrastructure and use of powerful data streaming and processing tools. To facilitate use of the dataset, we organize the tweet IDs we release by time (month and year) and language. This should enable interested researchers to work with the exact parts of the data related to their research questions even if they do not have large computing infrastructure. 

\noindent\textbf{Ethical Considerations.} We collect \texttt{Mega-COV} from the public domain (Twitter). In compliance with Twitter policy, we do not publish hydrated tweet content. Rather, we only publish publicly available tweet IDs. All Twitter policies, including respect and protection of user privacy, apply. We decided not to assign geographic region tags to the tweet IDs we distribute, but these already exist on the \texttt{json} object retrievable from Twitter. Still, location information should be used with caution. Twitter does not allow deriving or inferring, or storing derived or inferred, potentially sensitive characteristics about users. Sensitive user attributes identified by Twitter include health (e.g., pregnancy), negative financial status or condition, political affiliation or beliefs, religious or philosophical affiliation or beliefs, sex life or sexual orientation, trade union membership, and alleged or actual commission of a crime. If they decide to use \texttt{Mega-COV}, we expect researchers to review Twitter policy\footnote{ \url{https://developer.twitter.com/en/developer-terms/policy}} and applicable laws, including the European Union's General Data Protection Regulation (GDPR)\footnote{\url{https://gdpr-info.eu}.}, beforehand. \textit{\textbf{We encourage use of \texttt{Mega-COV} for social good, including applications that can improve health and well-being and enhance online safety.}} 
% \vspace*{-3mm}
\section{Related Works}\label{sec:rel}
%%%%%%%%%%%%%%%%%%%%%%%%%%%%%%%%%%%%%%
%%%%%%%%%%%%%%%%%%%%%%%%%%%%%%%%%%%%%%%%%%
\noindent\textbf{Twitter Datasets for COVID-19.}
%%%%%%%%%%%%%%%%%%%%%%%%%%%%%%%%%%%%%%%%%%
Several works have focused on creating datasets for enabling COVID-19 research. To the best of our knowledge, all these works depend on a list of hashtags related to COVID-19 and focus on a given period of time. For example,~\newcite{chen2020covid} started collecting tweets on Jan. $22^{nd}$ and continued updating by actively tracking a list of 22 popular keywords such as \textit{\#Coronavirus, \#Corona}, and \textit{\#Wuhancoronavirus}. As of May $30$, 2020~\cite{chen2020covid} report 144M tweets. %Our dataset is much larger, and does not have same distributional shift resulting from collection with hashtags.
~\newcite{singh2020first} collect a dataset covering January $16\ 2020$-March $15\ 2020$ using a list of hashtags such as \textit{\#2019nCoV, \#ChinaPneumonia} and \textit{\#ChinesePneumonia}, for a total of $2.8$M tweets, $\sim 18$M re-tweets, and $\sim 457$K direct conversations. Using location information on the data, authors report that tweets strongly correlated with newly identified cases in these locations. \\
Similarly,~\newcite{alqurashi2020large} use a list of keywords and hashtags related to Covid-19 with Twitter’s streaming API to collect a dataset of Arabic tweets. The dataset covers the period of March $1\ 2020$-March\ $30\ 2020$ and is at $4$M tweets. The authors' goal is to help researchers and policy makers study the various societal issues prevailing due to the pandemic. In the same vein, ~\newcite{lopez2020understanding} collect a dataset of $\sim 6.5$M in multiple languages, with English accounting for $\sim 63.4\%$ of the data. The dataset covers January $22\ 2020$-March $2020$. Analyzing the data, authors observe the level of retweets to rise abruptly as the crisis ramped up in Europe in late February and early March. \\
\textbf{Twitter in emergency and crisis.}
Social media can play a useful role in disaster and emergency since they provide a mechanism for wide information dissemination~\cite{simon2015socializing}. Examples include use of Twitter information for the Typhoon Haiyan in the Philippines~\cite{takahashi2015communicating}, Tsunami in Padang Indonesia~\cite{carley2016crowd}, the Nepal 2015 earthquakes~\cite{verma2019newswire}, Harvey Hurricane~\cite{marx2020sense}. A number of works have focused on developing systems for emergency response. An example is~\newcite{mccreadie2019trec}. Other works focused on developing systems for detecting misuse of social media~\cite{alshehri:2018:tacos,alshehri:2020:osact4,nagoudi-etal-2020-fake,elmadany:2020:leveraging}. \\
\textbf{Misinformation About COVID-19.}
Misinformation can spread fast during disaster. Social data have been used to study rumors and various types of fake information related to the Zika~\cite{ghenai2017catching} and Ebola~\cite{kalyanam2015facts} viruses. In the context of COVID-19, a number of works have focused on investigating the effect of misinformation on mental health~\cite{rosenberg2020twitter}, the types, sources, claims, and responses of a number of pieces of misinformation about COVID-19~\cite{brennen2020types}, the propagation pattern of rumours about COVID-19 on Twitter and Weibo~\cite{do2019rumour}, check-worthiness~\cite{wright2020fact}, modeling the spread of misinformation and related networks about the pandemic~\cite{cinelli2020covid,osho2020information,pierri2020topology,koubaa2020understanding}, estimating the rate of misinformation in COVID-19 associated tweets~\cite{kouzy2020coronavirus}, the use of bots~\cite{ferrara2020covid}, predicting whether a user is COVID19 positive or negative~\cite{karisani2020mining}, and the quality of shared links~\newcite{singh2020first}. Other works have focused on detecting racism and hate speech~\cite{devakumar2020racism,schild2020go,shimizu20202019,lyu2020sense} and emotional response~\cite{kleinberg2020measuring}.

\section{Conclusion}\label{sec:conc}
\noindent We presented \texttt{Mega-COV}, a billion-scale dataset of 104 languages for studying COVID-19 pandemic. In addition to being large and highly multilingual, our dataset comprises data pre-dating the pandemic. This allows for comparative and longitudinal investigations. We provided a global description of \texttt{Mega-COV} in terms of its geographic and temporal coverage, over-viewed its linguistic diversity, and provided analysis of its content based on hashtags and top domains. We also provided a case study of how the data can be used to track global human mobility. The scale of the Mega-COV has also allowed us to make a number of striking discoveries, including \textbf{(1)} the shift toward retweeting and replying to other users rather than tweeting in 2020 and \textbf{(2)} the role of international news sites as key sources of information during the pandemic. In addition, we developed effective models for detecting COVID relevance and COVID misinformation and applied them to a large sample of our dataset. Our dataset and models are publicly available.% for ethical research.  

\bibliography{eacl2021}

\begin{thebibliography}{47}
\expandafter\ifx\csname natexlab\endcsname\relax\def\natexlab#1{#1}\fi

\bibitem[{Abdul-Mageed and Ungar(2017)}]{mageed2017emonet}
Muhammad Abdul-Mageed and Lyle Ungar. 2017.
\newblock {EmoNet: Fine-grained emotion detection with gated recurrent neural
  networks}.
\newblock In \emph{Proceedings of the 55th Annual Meeting of the Association
  for Computational Linguistics (Volume 1: Long Papers)}, volume~1, pages
  718--728.

\bibitem[{Alqurashi et~al.(2020)Alqurashi, Alhindi, and
  Alanazi}]{alqurashi2020large}
Sarah Alqurashi, Ahmad Alhindi, and Eisa Alanazi. 2020.
\newblock Large arabic twitter dataset on covid-19.
\newblock \emph{arXiv preprint arXiv:2004.04315}.

\bibitem[{Alshehri et~al.(2020)Alshehri, Nagoudi, and
  Abdul-Mageed}]{alshehri:2020:osact4}
Ali Alshehri, El~Moatez~Billah Nagoudi, and Muhammad Abdul-Mageed. 2020.
\newblock Understanding and detecting dangerous speech in social media.
\newblock In \emph{Proceedings of the 4th Workshop on Open-Source Arabic
  Corpora and Processing Tools, with a Shared Task on Offensive Language
  Detection}, pages 40--47, Marseille, France. European Language Resource
  Association.

\bibitem[{Alshehri et~al.(2018)Alshehri, Nagoudi, Hassan, and
  Abdul-Mageed}]{alshehri:2018:tacos}
Ali Alshehri, El~Moatez~Billah Nagoudi, Alhuzali Hassan, and Muhammad
  Abdul-Mageed. 2018.
\newblock {Think before your click: Data and models for adult content in Arabic
  twitter}.
\newblock In \emph{The 2nd Text Analytics for Cybersecurity and Online Safety
  (TA-COS-2018), LREC}.

\bibitem[{Brennen et~al.(2020)Brennen, Simon, Howard, and
  Nielsen}]{brennen2020types}
J~Scott Brennen, Felix~M Simon, Philip~N Howard, and Rasmus~Kleis Nielsen.
  2020.
\newblock Types, sources, and claims of covid-19 misinformation.
\newblock \emph{Reuters Institute}.

\bibitem[{Carley et~al.(2016)Carley, Malik, Landwehr, Pfeffer, and
  Kowalchuck}]{carley2016crowd}
Kathleen~M Carley, Momin Malik, Peter~M Landwehr, J{\"u}rgen Pfeffer, and
  Michael Kowalchuck. 2016.
\newblock Crowd sourcing disaster management: The complex nature of twitter
  usage in padang indonesia.
\newblock \emph{Safety science}, 90:48--61.

\bibitem[{Chen et~al.(2020)Chen, Lerman, and Ferrara}]{chen2020covid}
Emily Chen, Kristina Lerman, and Emilio Ferrara. 2020.
\newblock Covid-19: The first public coronavirus twitter dataset.
\newblock \emph{arXiv preprint arXiv:2003.07372}.

\bibitem[{Cinelli et~al.(2020)Cinelli, Quattrociocchi, Galeazzi, Valensise,
  Brugnoli, Schmidt, Zola, Zollo, and Scala}]{cinelli2020covid}
Matteo Cinelli, Walter Quattrociocchi, Alessandro Galeazzi, Carlo~Michele
  Valensise, Emanuele Brugnoli, Ana~Lucia Schmidt, Paola Zola, Fabiana Zollo,
  and Antonio Scala. 2020.
\newblock The covid-19 social media infodemic.
\newblock \emph{arXiv preprint arXiv:2003.05004}.

\bibitem[{Conneau et~al.(2020)Conneau, Khandelwal, Goyal, Chaudhary, Wenzek,
  Guzm{\'a}n, Grave, Ott, Zettlemoyer, and
  Stoyanov}]{conneau-etal-2020-unsupervised}
Alexis Conneau, Kartikay Khandelwal, Naman Goyal, Vishrav Chaudhary, Guillaume
  Wenzek, Francisco Guzm{\'a}n, Edouard Grave, Myle Ott, Luke Zettlemoyer, and
  Veselin Stoyanov. 2020.
\newblock \href {https://doi.org/10.18653/v1/2020.acl-main.747} {Unsupervised
  cross-lingual representation learning at scale}.
\newblock In \emph{Proceedings of the 58th Annual Meeting of the Association
  for Computational Linguistics}, pages 8440--8451, Online. Association for
  Computational Linguistics.

\bibitem[{Cui and Lee(2020)}]{cui2020coaid}
Limeng Cui and Dongwon Lee. 2020.
\newblock Coaid: Covid-19 healthcare misinformation dataset.
\newblock \emph{arXiv preprint arXiv:2006.00885}.

\bibitem[{Devakumar et~al.(2020)Devakumar, Shannon, Bhopal, and
  Abubakar}]{devakumar2020racism}
Delan Devakumar, Geordan Shannon, Sunil~S Bhopal, and Ibrahim Abubakar. 2020.
\newblock Racism and discrimination in covid-19 responses.
\newblock \emph{Lancet (London, England)}, 395(10231):1194.

\bibitem[{Devlin et~al.(2018)Devlin, Chang, Lee, and
  Toutanova}]{devlin2018bert}
Jacob Devlin, Ming-Wei Chang, Kenton Lee, and Kristina Toutanova. 2018.
\newblock Bert: Pre-training of deep bidirectional transformers for language
  understanding.
\newblock \emph{arXiv preprint arXiv:1810.04805}.

\bibitem[{Do et~al.(2019)Do, Luo, Nguyen, and Deligiannis}]{do2019rumour}
Tien~Huu Do, Xiao Luo, Duc~Minh Nguyen, and Nikos Deligiannis. 2019.
\newblock Rumour detection via news propagation dynamics and user
  representation learning.
\newblock \emph{arXiv preprint arXiv:1905.03042}.

\bibitem[{Do et~al.(2018)Do, Nguyen, Tsiligianni, Cornelis, and
  Deligiannis}]{do2018twitter}
Tien~Huu Do, Duc~Minh Nguyen, Evaggelia Tsiligianni, Bruno Cornelis, and Nikos
  Deligiannis. 2018.
\newblock Twitter user geolocation using deep multiview learning.
\newblock In \emph{2018 IEEE International Conference on Acoustics, Speech and
  Signal Processing (ICASSP)}, pages 6304--6308. IEEE.

\bibitem[{Elmadany et~al.(2020)Elmadany, Zhang, Abdul-Mageed, and
  Hashemi}]{elmadany:2020:leveraging}
AbdelRahim Elmadany, Chiyu Zhang, Muhammad Abdul-Mageed, and Azadeh Hashemi.
  2020.
\newblock Leveraging affective bidirectional transformers for offensive
  language detection.
\newblock In \emph{The 4th Workshop on Open-Source Arabic Corpora and
  Processing Tools (OSACT4), LREC}, pages 102--108.

\bibitem[{Ferrara(2020)}]{ferrara2020covid}
Emilio Ferrara. 2020.
\newblock \# covid-19 on twitter: Bots, conspiracies, and social media
  activism.
\newblock \emph{arXiv preprint arXiv:2004.09531}.

\bibitem[{Ghenai and Mejova(2017)}]{ghenai2017catching}
Amira Ghenai and Yelena Mejova. 2017.
\newblock Catching zika fever: Application of crowdsourcing and machine
  learning for tracking health misinformation on twitter.
\newblock \emph{arXiv preprint arXiv:1707.03778}.

\bibitem[{Go et~al.(2009)Go, Bhayani, and Huang}]{go2009twitter}
Alec Go, Richa Bhayani, and Lei Huang. 2009.
\newblock Twitter sentiment classification using distant supervision.
\newblock \emph{CS224N project report, Stanford}, 1(12):2009.

\bibitem[{Graham et~al.(2014)Graham, Hale, and Gaffney}]{graham2014world}
Mark Graham, Scott~A Hale, and Devin Gaffney. 2014.
\newblock Where in the world are you? geolocation and language identification
  in twitter.
\newblock \emph{The Professional Geographer}, 66(4):568--578.

\bibitem[{Han et~al.(2016)Han, Rahimi, Derczynski, and
  Baldwin}]{han2016twitter}
Bo~Han, Afshin Rahimi, Leon Derczynski, and Timothy Baldwin. 2016.
\newblock Twitter geolocation prediction shared task of the 2016 workshop on
  noisy user-generated text.
\newblock In \emph{Proceedings of the 2nd Workshop on Noisy User-generated Text
  (WNUT)}, pages 213--217.

\bibitem[{Kalyanam et~al.(2015)Kalyanam, Velupillai, Doan, Conway, and
  Lanckriet}]{kalyanam2015facts}
Janani Kalyanam, Sumithra Velupillai, Son Doan, Mike Conway, and Gert
  Lanckriet. 2015.
\newblock Facts and fabrications about ebola: A twitter based study.
\newblock \emph{arXiv preprint arXiv:1508.02079}.

\bibitem[{Karisani and Karisani(2020)}]{karisani2020mining}
Negin Karisani and Payam Karisani. 2020.
\newblock Mining coronavirus (covid-19) posts in social media.
\newblock \emph{arXiv preprint arXiv:2004.06778}.

\bibitem[{Kleinberg et~al.(2020)Kleinberg, van~der Vegt, and
  Mozes}]{kleinberg2020measuring}
Bennett Kleinberg, Isabelle van~der Vegt, and Maximilian Mozes. 2020.
\newblock Measuring emotions in the covid-19 real world worry dataset.
\newblock \emph{arXiv preprint arXiv:2004.04225}.

\bibitem[{Koubaa(2020)}]{koubaa2020understanding}
Anis Koubaa. 2020.
\newblock Understanding the covid19 outbreak: A comparative data analytics and
  study.
\newblock \emph{arXiv preprint arXiv:2003.14150}.

\bibitem[{Kouzy et~al.(2020)Kouzy, Abi~Jaoude, Kraitem, El~Alam, Karam, Adib,
  Zarka, Traboulsi, Akl, and Baddour}]{kouzy2020coronavirus}
Ramez Kouzy, Joseph Abi~Jaoude, Afif Kraitem, Molly~B El~Alam, Basil Karam,
  Elio Adib, Jabra Zarka, Cindy Traboulsi, Elie~W Akl, and Khalil Baddour.
  2020.
\newblock Coronavirus goes viral: Quantifying the covid-19 misinformation
  epidemic on twitter.
\newblock \emph{Cureus}, 12(3).

\bibitem[{Lopez et~al.(2020)Lopez, Vasu, and
  Gallemore}]{lopez2020understanding}
Christian~E Lopez, Malolan Vasu, and Caleb Gallemore. 2020.
\newblock Understanding the perception of covid-19 policies by mining a
  multilanguage twitter dataset.
\newblock \emph{arXiv preprint arXiv:2003.10359}.

\bibitem[{Lui and Baldwin(2012)}]{lui2012langid}
Marco Lui and Timothy Baldwin. 2012.
\newblock langid. py: An off-the-shelf language identification tool.
\newblock In \emph{Proceedings of the ACL 2012 system demonstrations}, pages
  25--30. Association for Computational Linguistics.

\bibitem[{Lui and Baldwin(2014)}]{lui2014accurate}
Marco Lui and Timothy Baldwin. 2014.
\newblock Accurate language identification of twitter messages.
\newblock In \emph{Proceedings of the 5th workshop on language analysis for
  social media (LASM)}, pages 17--25.

\bibitem[{Lyu et~al.(2020)Lyu, Chen, Wang, and Luo}]{lyu2020sense}
Hanjia Lyu, Long Chen, Yu~Wang, and Jiebo Luo. 2020.
\newblock Sense and sensibility: Characterizing social media users regarding
  the use of controversial terms for covid-19.
\newblock \emph{arXiv preprint arXiv:2004.06307}.

\bibitem[{Marx et~al.(2020)Marx, Mirbabaie, and Ehnis}]{marx2020sense}
Julian Marx, Milad Mirbabaie, and Christian Ehnis. 2020.
\newblock Sense-giving strategies of media organisations in social media
  disaster communication: Findings from hurricane harvey.
\newblock \emph{arXiv preprint arXiv:2004.08567}.

\bibitem[{McCreadie et~al.(2019)McCreadie, Buntain, and
  Soboroff}]{mccreadie2019trec}
Richard McCreadie, Cody Buntain, and Ian Soboroff. 2019.
\newblock Trec incident streams: Finding actionable information on social
  media.
\newblock \emph{Proceedings of the 16th International Conferenc e on
  Information Systems for Crisis Response and Management, Valencia, Spain}.

\bibitem[{Mohammad and Kiritchenko(2015)}]{mohammad2015using}
Saif~M Mohammad and Svetlana Kiritchenko. 2015.
\newblock Using hashtags to capture fine emotion categories from tweets.
\newblock \emph{Computational Intelligence}, 31(2):301--326.

\bibitem[{Nagoudi et~al.(2020)Nagoudi, Elmadany, Abdul-Mageed, Alhindi, and
  Cavusoglu}]{nagoudi-etal-2020-fake}
El~Moatez~Billah Nagoudi, AbdelRahim Elmadany, Muhammad Abdul-Mageed, Tariq
  Alhindi, and Hasan Cavusoglu. 2020.
\newblock {Machine Generation and Detection of Arabic Manipulated and Fake
  News}.
\newblock In \emph{Proceedings of the Fifth Arabic Natural Language Processing
  Workshop}, pages 69--84.

\bibitem[{Ooms and Sites(2018)}]{ooms2018cld2}
J~Ooms and D~Sites. 2018.
\newblock {cld2: Google’s Compact Language Detector 2}.
\newblock \emph{Retrieved Feburary}, 7:2019.

\bibitem[{Osho et~al.(2020)Osho, Waters, and Amariucai}]{osho2020information}
Abiola Osho, Caden Waters, and George Amariucai. 2020.
\newblock An information diffusion approach to rumor propagation and
  identification on twitter.
\newblock \emph{arXiv preprint arXiv:2002.11104}.

\bibitem[{Pierri et~al.(2020)Pierri, Piccardi, and Ceri}]{pierri2020topology}
Francesco Pierri, Carlo Piccardi, and Stefano Ceri. 2020.
\newblock Topology comparison of twitter diffusion networks effectively reveals
  misleading information.
\newblock \emph{Scientific reports}, 10(1):1--9.

\bibitem[{Roller et~al.(2012)Roller, Speriosu, Rallapalli, Wing, and
  Baldridge}]{roller2012supervised}
Stephen Roller, Michael Speriosu, Sarat Rallapalli, Benjamin Wing, and Jason
  Baldridge. 2012.
\newblock Supervised text-based geolocation using language models on an
  adaptive grid.
\newblock In \emph{Proceedings of the 2012 Joint Conference on Empirical
  Methods in Natural Language Processing and Computational Natural Language
  Learning}, pages 1500--1510. Association for Computational Linguistics.

\bibitem[{Rosenberg et~al.(2020)Rosenberg, Syed, and
  Rezaie}]{rosenberg2020twitter}
Hans Rosenberg, Shahbaz Syed, and Salim Rezaie. 2020.
\newblock The twitter pandemic: The critical role of twitter in the
  dissemination of medical information and misinformation during the covid-19
  pandemic.
\newblock \emph{Canadian Journal of Emergency Medicine}, pages 1--7.

\bibitem[{Schild et~al.(2020)Schild, Ling, Blackburn, Stringhini, Zhang, and
  Zannettou}]{schild2020go}
Leonard Schild, Chen Ling, Jeremy Blackburn, Gianluca Stringhini, Yang Zhang,
  and Savvas Zannettou. 2020.
\newblock " go eat a bat, chang!": An early look on the emergence of sinophobic
  behavior on web communities in the face of covid-19.
\newblock \emph{arXiv preprint arXiv:2004.04046}.

\bibitem[{Shimizu(2020)}]{shimizu20202019}
Kazuki Shimizu. 2020.
\newblock 2019-ncov, fake news, and racism.
\newblock \emph{The Lancet}, 395(10225):685--686.

\bibitem[{Simon et~al.(2015)Simon, Goldberg, and Adini}]{simon2015socializing}
Tomer Simon, Avishay Goldberg, and Bruria Adini. 2015.
\newblock Socializing in emergencies—a review of the use of social media in
  emergency situations.
\newblock \emph{International Journal of Information Management},
  35(5):609--619.

\bibitem[{Singh et~al.(2020)Singh, Bansal, Bode, Budak, Chi, Kawintiranon,
  Padden, Vanarsdall, Vraga, and Wang}]{singh2020first}
Lisa Singh, Shweta Bansal, Leticia Bode, Ceren Budak, Guangqing Chi, Kornraphop
  Kawintiranon, Colton Padden, Rebecca Vanarsdall, Emily Vraga, and Yanchen
  Wang. 2020.
\newblock A first look at covid-19 information and misinformation sharing on
  twitter.
\newblock \emph{arXiv preprint arXiv:2003.13907}.

\bibitem[{Takahashi et~al.(2015)Takahashi, Tandoc~Jr, and
  Carmichael}]{takahashi2015communicating}
Bruno Takahashi, Edson~C Tandoc~Jr, and Christine Carmichael. 2015.
\newblock Communicating on twitter during a disaster: An analysis of tweets
  during typhoon haiyan in the philippines.
\newblock \emph{Computers in Human Behavior}, 50:392--398.

\bibitem[{Verma et~al.(2019)Verma, Karimi, Lee, Gnawali, and
  Shakery}]{verma2019newswire}
Rakesh Verma, Samaneh Karimi, Daniel Lee, Omprakash Gnawali, and Azadeh
  Shakery. 2019.
\newblock Newswire versus social media for disaster response and recovery.
\newblock \emph{arXiv preprint arXiv:1906.10607}.

\bibitem[{WHO(2020)}]{world2020statement}
WHO. 2020.
\newblock Who statement regarding cluster of pneumonia cases in wuhan, china.
\newblock \emph{Beijing: WHO}, 9.

\bibitem[{Wright and Augenstein(2020)}]{wright2020fact}
Dustin Wright and Isabelle Augenstein. 2020.
\newblock Fact check-worthiness detection as positive unlabelled learning.
\newblock \emph{arXiv preprint arXiv:2003.02736}.

\bibitem[{Zhou et~al.(2020)Zhou, Mulay, Ferrara, and
  Zafarani}]{zhou2020recovery}
Xinyi Zhou, Apurva Mulay, Emilio Ferrara, and Reza Zafarani. 2020.
\newblock Recovery: A multimodal repository for covid-19 news credibility
  research.
\newblock \emph{arXiv preprint arXiv:2006.05557}.

\end{thebibliography}
\bibliographystyle{acl_natbib}

% \newpage
\appendix
\clearpage
\twocolumn[{%
 \centering
 \Large \textbf{}\\[5em]

}]
\appendixpage
\addappheadtotoc
\counterwithin{figure}{section}
\counterwithin{table}{section}

\section{Data Distribution}\label{sec:append:data_geo}
%-----------------

%-------------
\begin{figure}[ht]
\tiny  
 \hspace*{-0.25in}
  \includegraphics[width=9cm]{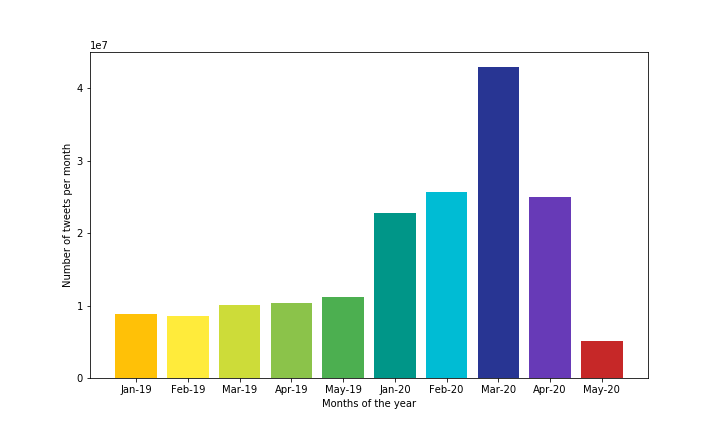}
 
  \caption{Frequency of tweeting during Jan-May (10$^{th}$) 2020 vs. Jan-May 2019.}
\label{fig:avg-tweeting}
\end{figure}

%%%%%%%%%%%%%%%%%%%%%%%%%%%%%%
\begin{figure}[]
  \centering
  \frame{\includegraphics[width=7cm]{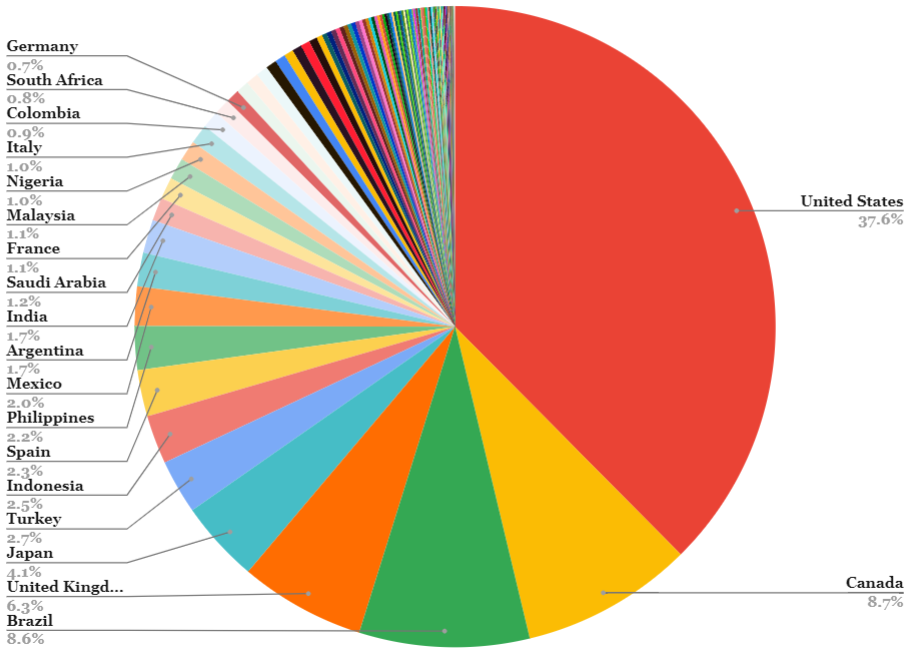}}
  \caption{Geographical diversity in \texttt{Mega-COV}. We show the distribution of our \textit{geo-located} data over the top 20 countries with most tweets and responses. Overall, $268$ countries are represented in the data.}
\label{fig:geo-diversity}
\end{figure}

%----------------
%%%%%%%%%%%%%%%%%%%
\section{Hashtag Content Analysis}\label{sec:append:hashtags}
%%%%%%%%%%%%%%%%%%%

%----

%%%%%%%%%%
\begin{table}[h!]
\footnotesize 
\centering
\begin{tabular}{ll|ll}
\hline
%\multicolumn{4}{c}{All Data} \\
%\hline
\multicolumn{2}{c}{\textbf{2019}}  & \multicolumn{2}{c}{\textbf{2020}} \\
\hline
\multicolumn{1}{c}{\textbf{Hashtag}}             & \multicolumn{1}{c}{\textbf{Freq}} & \multicolumn{1}{c}{\textbf{Hashtag}}             & \multicolumn{1}{c}{\textbf{Freq}} \\
\hline
NewProfilePic        & 64,922     & COVID19          & 260,024    \\
love                 & 41,964     & coronavirus      & 219,615    \\
Repost               & 39,128     & NewProfilePic    & 102,724    \\
art                  & 35,825     & BBB20            & 91,775     \\
% photography          & 34,874     & \foreignlanguage{japanese}{質問箱}              & 86,220     \\
music                & 28,335     & Covid-19        & 70,106     \\
travel               & 28,236     & COVID-19         & 67,737     \\
% fashion              & 23,814     & \foreignlanguage{japanese}{匿名質問募集中}          & 56,345     \\
GameofThrones        & 21,484     & Coronavirus      & 53,251     \\
nature               & 18,563     & covid19          & 47,940     \\
instagood            & 18,491     & StayHome         & 44,165     \\
photooftheday        & 18,032     & NintendoSwitch   & 42,812     \\
tbt                  & 17,332     & love             & 42,497     \\
realestate           & 17,255     & bbb20            & 39,974     \\
shopmycloset         & 16,760     & NowPlaying       & 39,069     \\
GameOfThrones        & 16,127     & Repost           & 37,036     \\
peing                & 15,930     & AnimalCrossing   & 36,369     \\
% \foreignlanguage{japanese}{質問箱}                  & 15,803     & art              & 35,570     \\
fitness              & 15,623     & ACNH             & 35,528     \\
food                 & 15,358     & photography      & 35,209     \\
BellLetsTalk         & 14,853     & COVID2019        & 33,512     \\
NowPlaying           & 14,849     & shopmycloset     & 31,428     \\
family               & 14,060     & music            & 30,537     \\
style                & 14,041     & StayAtHome       & 30,313     \\
SoundCloud           & 13,904     & QuedateEnCasa    & 30,194     \\
WeTheNorth           & 13,579     & stayhome         & 27,540     \\
GOT                  & 13,458     & PS4share         & 27,487     \\
np                   & 13,335     & SocialDistancing & 27,376     \\
MyTwitterAnniv. & 12,965     & lockdown         & 27,344     \\
Toronto              & 12,964     & TikTok           & 27,287        \\
\hline
\end{tabular}%
\caption{Top 30 hashtags in \texttt{Mega-COV} for 2019 vs. 2020.}
\label{tab:ht_all}
\end{table}
%%%%%%%%%%%%%%%%%%
%----------------

\begin{figure*}[h]

\includegraphics[width=\linewidth]{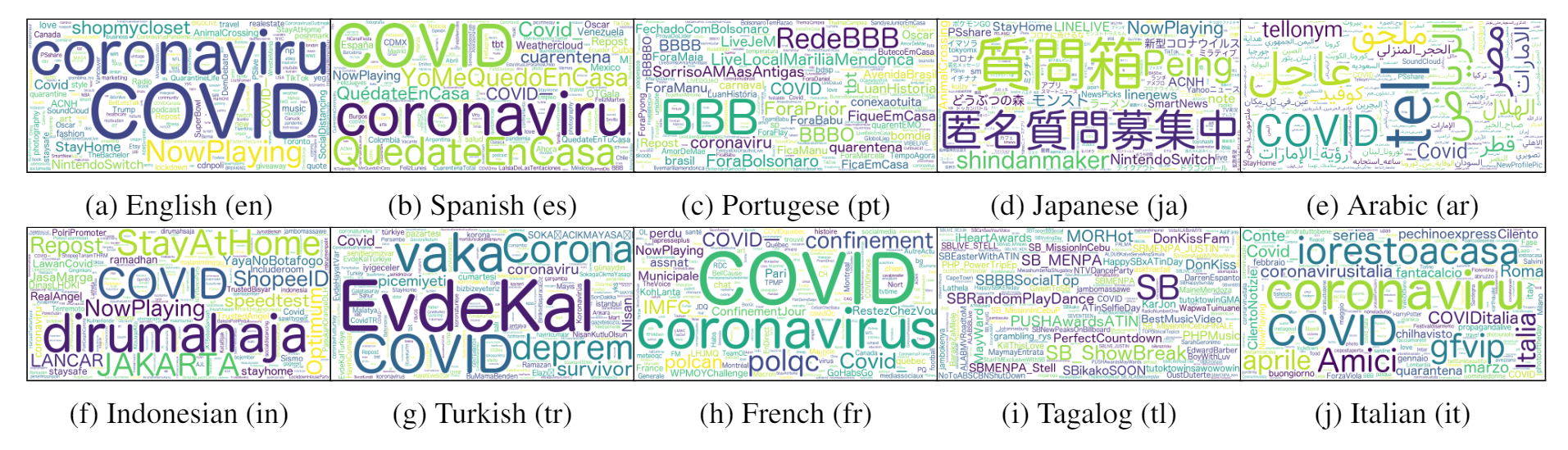}

\caption{Word clouds for hashtags in tweets from the top 10 languages in the data. We note that tweets in non-English can still carry English hashtags or employ Latin script.}
\label{fig:wcs}
\end{figure*}

 %--------------------
 
%------------
\noindent Hashtags usually correlate with the topics users post about. We provide the top 30 hashtags in the data in Table~\ref{tab:ht_all}. As the table shows, users tweet heavily about the pandemic using hashtags such as \textit{COVID19, coronavirus, Coronavirus, COVID\-19, Covid\-19, covid19} and \textit{StayAtHome}. Simple word clouds of hashtags from the various languages (Figure~\ref{fig:wcs} provides clouds from the top 10 languages) also show COVID-19 topics trending. We also observe hashtags related to \textbf{gaming} (e.g.,  \textit{NowPlaydo, PSshare}, and \textit{NintendoSwitch}). This reflects how users may be spending part of their newly-found home time. We also note frequent occurrence of \textbf{political} hashtags in languages such Arabic, Farsi, Indian, and Urdu. This is in contrast to discussions in European languages where politics are not as visible. For example, in Urdu, discussions involving the army and border issues show up. This may be partly due to different political environments, but also due to certain European countries such as Italy, Sweden, Spain, and the U.K. being hit harder (and earlier) than many countries in the Middle East and Asia. In Indian languages such as Tamil and Hindi, posts also focused on movies (e.g., \textit{Valimai}), TV shows (e.g., \textit{Big Boss}), doctors, and even fake news along with the pandemic-related hashtags.

An interesting observation from the \textbf{Chinese language} word cloud is the use of hashtags such as \textit{ChinaPneumonia} and \textit{WuhanPneumonia} to refer to the pandemic. We did not observe these same hashtags in any of the other languages. Additionally, for some reason, \textbf{Apple} seems to be trending during the first 4 months of 2020 in China owing to hashtags such as \textit{appledaily} and \textit{appledailytw}. Some languages, such as Romanian and Vietnamese, involve discussions of \textbf{bitcoin} and crypto-currency. This was also seen in the Chinese language word cloud, but not as prominently. %Another surprising observation is seen from the Finnish language where users post about the corona virus and gaming but also about \textit{kirtan}, \textit{gurbani} which are \textbf{religious} terms related to Sikh religion.

\section{Domain Sharing Analysis}\label{sec:append:domains}

\noindent Domains in URLs shared by users also provide a window on what is share-worthy. We perform an analysis of the top 200 domains shared in each of 2019 and 2020. The major observation we reach is the \textit{surge in tweets involving news websites, and the rise in ranks for the majority of these websites compared to 2019}. Table~\ref{tab:news} shows the top 40 news domains in the 2020 data and their change in rank compared to 2019. Such a heavy sharing of news domains reflects users' needs: Intuitively, at times of global disruption, people need more frequent updates on ongoing events. Of particular importance, especially relative to other ongoing political polarization in the U.S., is the striking rise of the conservative news network \textit{Fox News}, which has moved from a rank of 118 in 2019 to 67 in 2020 with a swooping 51 positions jump. We also note the rank of some news sites (e.g., \textit{The Globe and Mail} and \textit{The Star} going down. This is perhaps due to people resorting to international (and more diverse) sources of information to remain informed about countries other than their own.\\
\textbf{Other domains}: Other noteworthy domain activities include those related to gaming, video and music, and social media tools. Ranks of these domains have not necessarily shifted higher than 2019 but remain prominent. This shows these themes still being relevant in 2020. In spite of the economic impact of the pandemic, shopping domains such as \textit{etsy.me} and \textit{poshmark.com} have markedly risen in rank as people moved to shopping online in more significant ways. We now introduce a case study as to how our data can be used for mobility tracking.

\begin{table}[]
\begin{adjustbox}{width=\columnwidth}
\begin{tabular}{lr|lr}
\hline
\textbf{Domain}             & \textbf{Rank} & \textbf{Domain}              & \textbf{Rank}  \\ \hline
theguardian.com    & $\uparrow$    3           & thehill.com         & $\uparrow$    51          \\ 
nytimes.com        & $\uparrow$    10          & globeandmail.com & $\downarrow$    -38         \\ 
cnn.com            & $\uparrow$    18          & businessinsdr.com & $\uparrow$    31          \\ 
apple.news         & $\uparrow$    4           & theatlantic.com     & $\uparrow$    27          \\ 
washingtonpost.com & $\uparrow$    16          & newsbreakapp.com    & $\uparrow$    472         \\ 
cbc.ca             &  $\downarrow$    -13         & eldiario.es         & $\uparrow$    62          \\ 
bbc.co.uk          & $\downarrow$    -4          & apnews.com          & $\uparrow$    48          \\ 
bbc.com            & $\uparrow$    3           & abc.es              & $\uparrow$    89          \\ 
nyti.ms            & $\downarrow$    -11         & reuters.com         & $\uparrow$    59          \\ 
foxnews.com        & $\uparrow$    51          & thestar.com         & $\downarrow$    -64         \\ 
forbes.com         & $\downarrow$    -14         & francebleu.fr       & $\uparrow$    424         \\ 
nbcnews.com        & $\uparrow$    39          & globalnews.ca       & $\downarrow$    -78         \\ 
wsj.com            & $\uparrow$    11          & independent.co.uk   & $\downarrow$    -10         \\ 
bloomberg.com      & $\uparrow$    13          & elmundo.es          & $\uparrow$    21          \\ 
ctvnews.ca         & $\uparrow$    2           & indiatimes.com      &$\updownarrow$    0           \\
nypost.com         & $\uparrow$    100         & radio-canada.ca     & $\downarrow$    -66         \\ 
cnbc.com           & $\uparrow$    43          & lavanguardia.com    & $\uparrow$    96          \\ 
usatoday.com       & $\uparrow$    6           & dailymail.co.uk     & $\uparrow$    23          \\ 
latimes.com        & $\uparrow$    23          & politico.com        & $\uparrow$    403         \\ 
huffpost.com       & $\uparrow$    66          & sky.com             & $\uparrow$    114         \\ \hline
\end{tabular}%
\end{adjustbox}
 \caption{Top 40 domains in 2020 data and their rank change relative to their rank in 2019.}
 \label{tab:news}
\end{table}

%%%%%%%%%%%%%%%%%%%%%%%%%%%%%%%%%
\section{Case Study: Mapping Human Mobility with Mega-COV}\label{sec:hum-mob}
%%%%%%%%%%%%%%%%%%%%%%%%%%%%%%%%%
\noindent Geolocation information in \texttt{Mega-COV} can be used to characterize and track human mobility in various ways. We investigate some of these next.

% %-----------
\begin{figure*}[t]
\center
\begin{subfigure}{.32\linewidth}
  \centering
  \frame{\includegraphics[width=\linewidth]{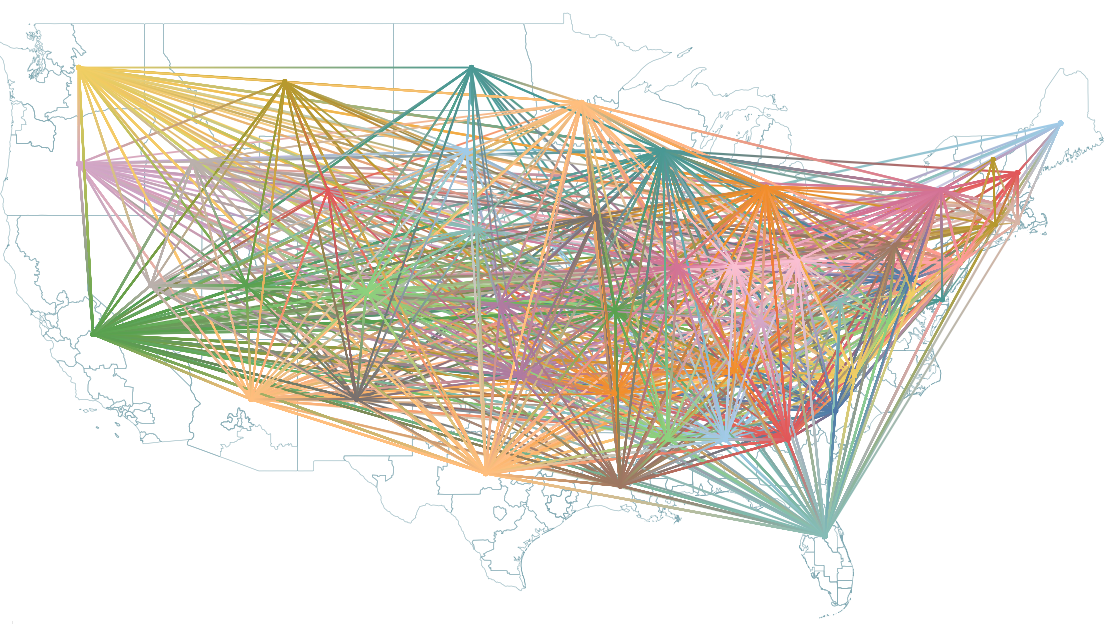}}
  \caption{Overall inter-state mobility}
  \label{fig:us-mobility_a}
\end{subfigure}
\begin{subfigure}{.32\linewidth}
  \centering
  \frame{\includegraphics[width=\linewidth]{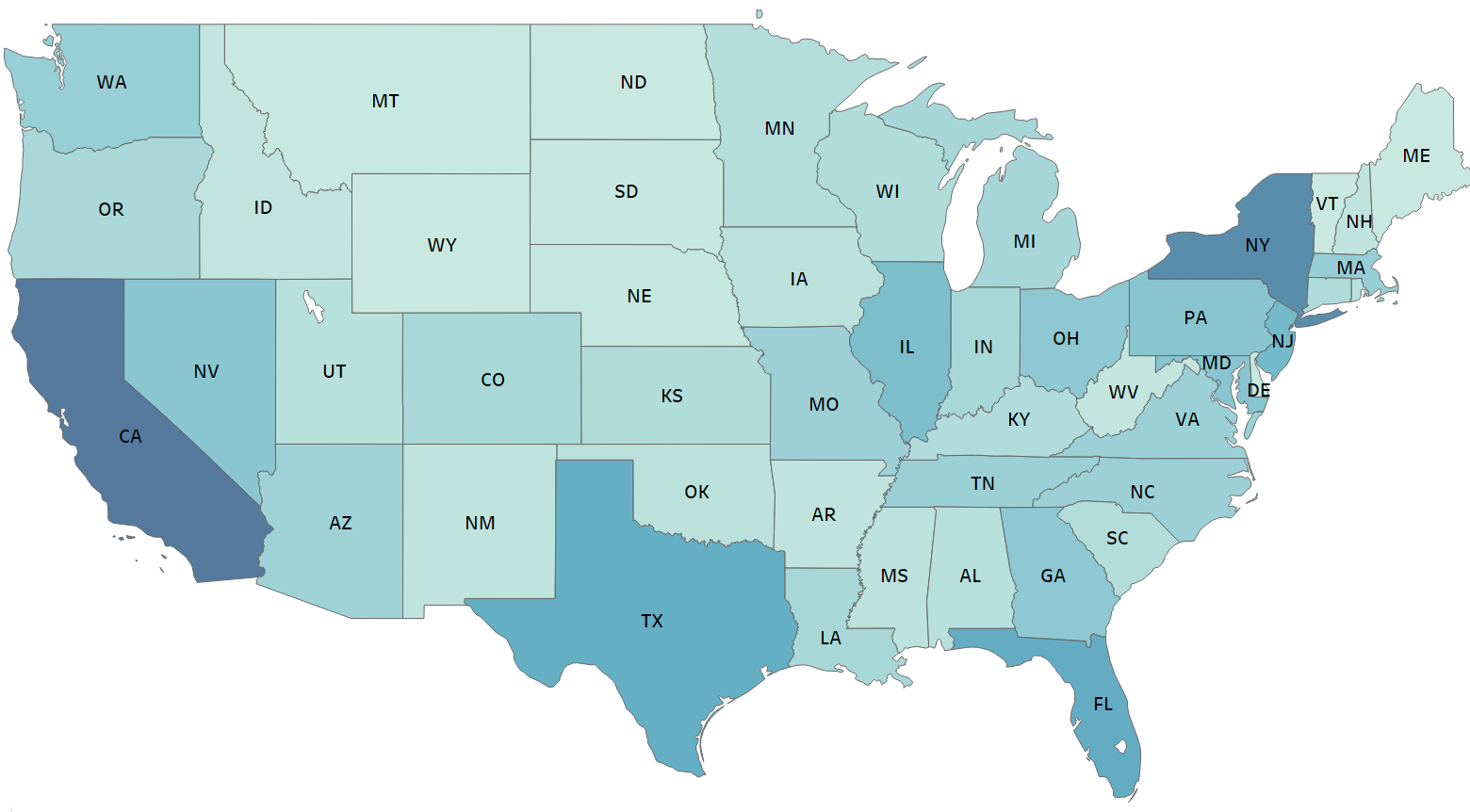}}
  \caption{January}
  \label{fig:us-mobility_b}
\end{subfigure}
\begin{subfigure}{.32\linewidth}
  \centering
  \frame{\includegraphics[width=\linewidth]{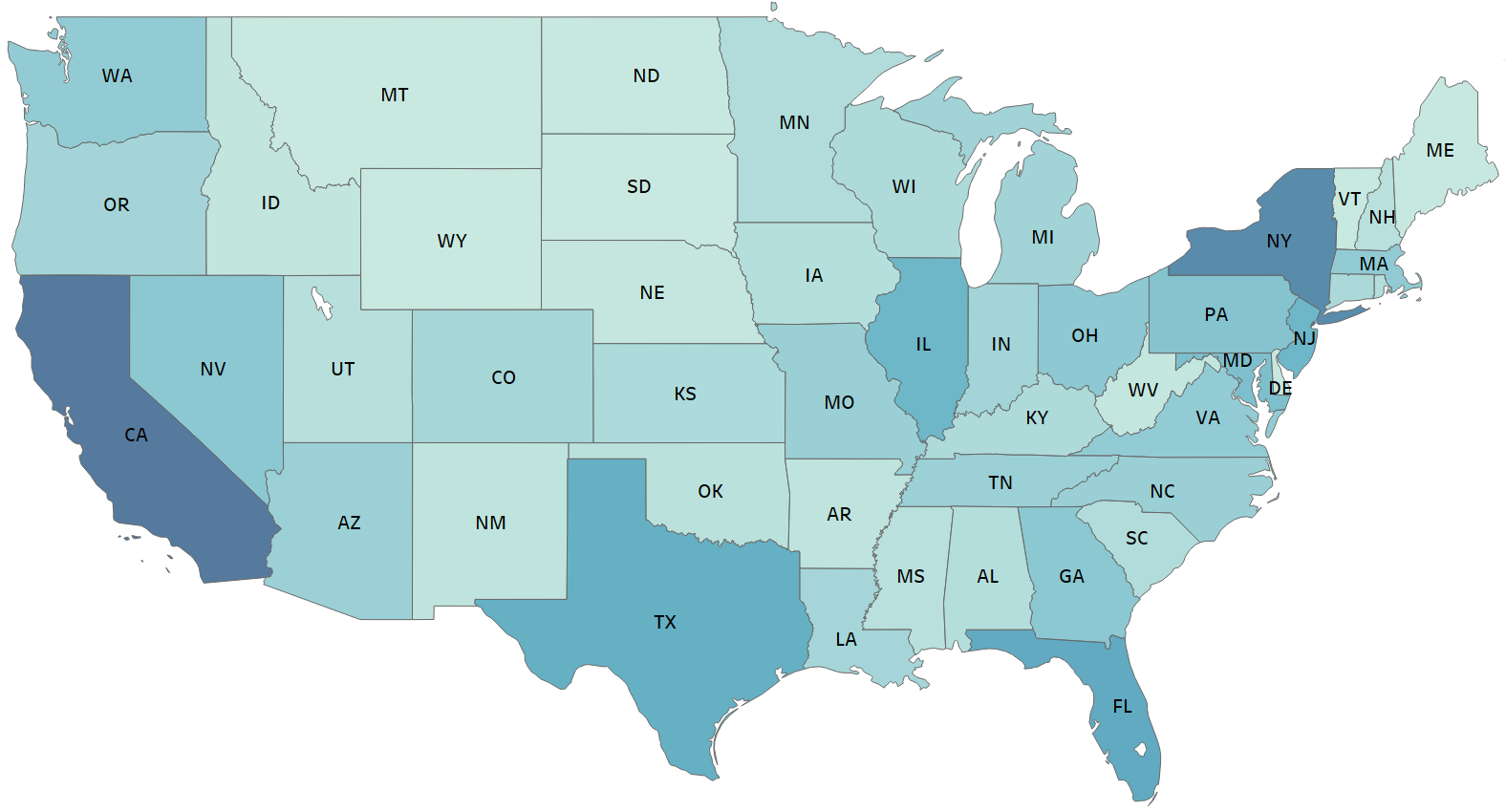}}
  \caption{February}
  \label{fig:us-mobility_c}
\end{subfigure}\\
\begin{subfigure}{.32\linewidth}
  \centering
  \frame{\includegraphics[width=\linewidth]{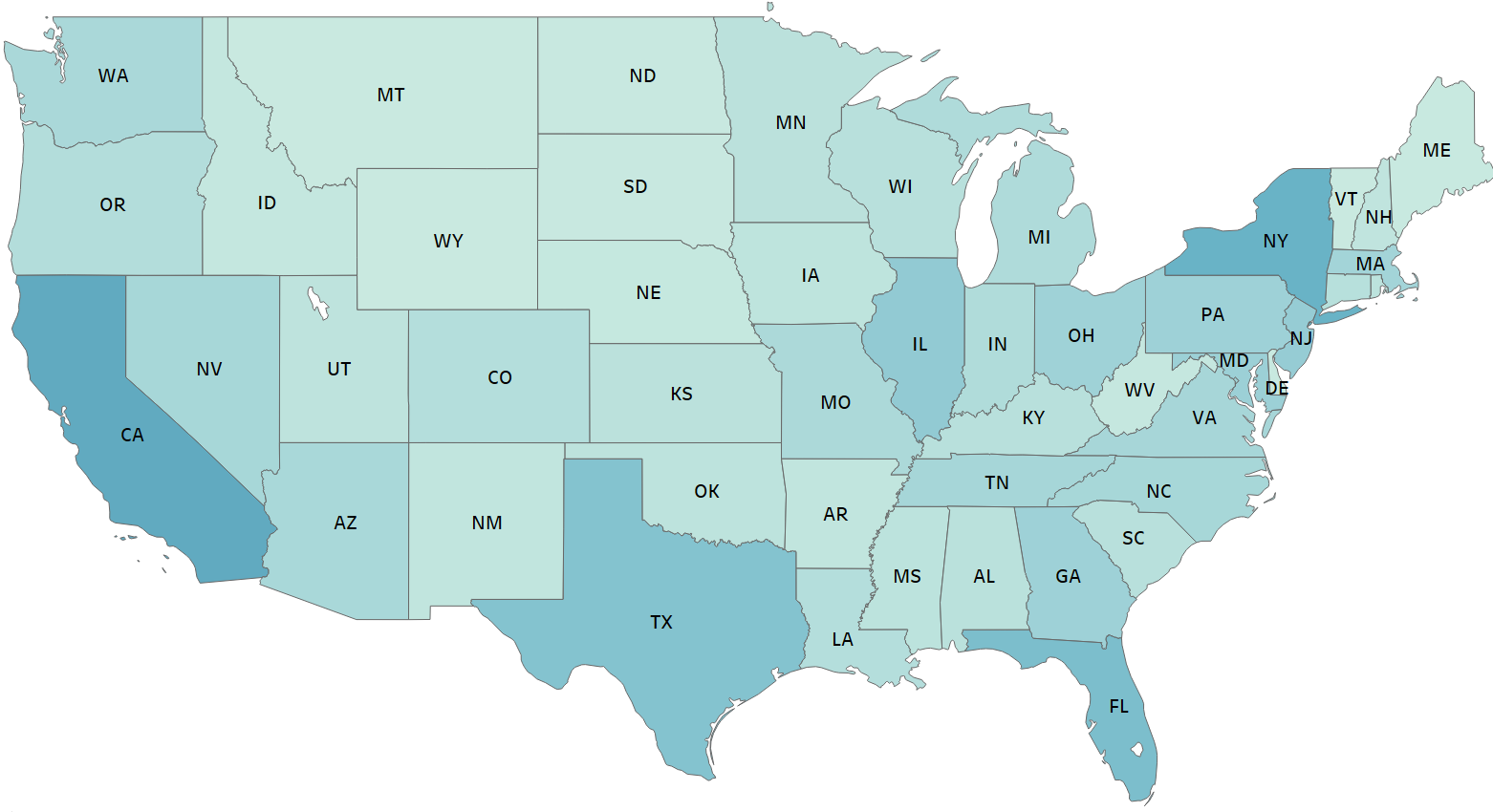}}
  \caption{March}
  \label{fig:sfigus-mobility_d}
\end{subfigure}
\begin{subfigure}{.33\linewidth}
  \centering
  \frame{\includegraphics[width=\linewidth]{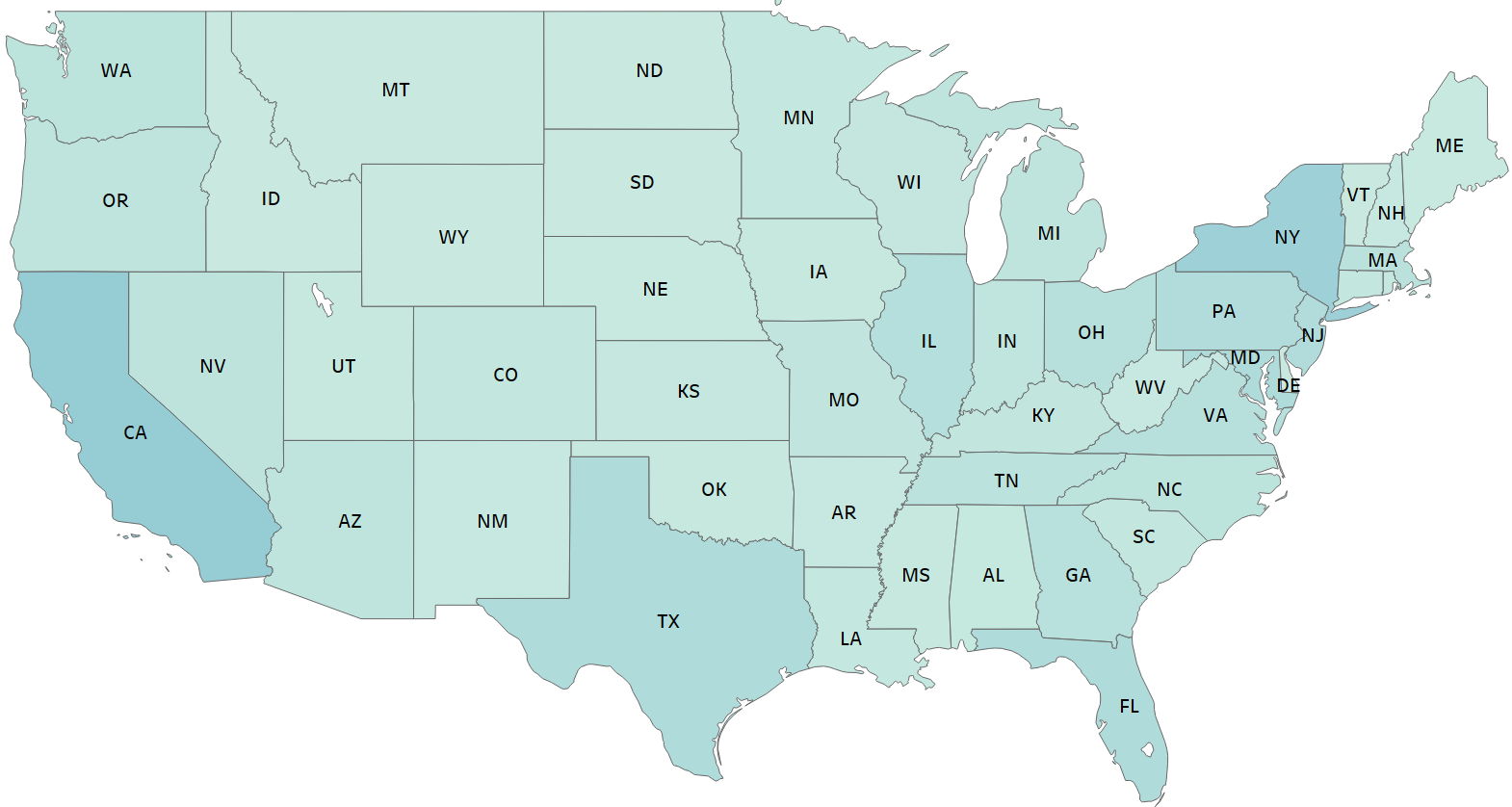}}
  \caption{April}
  \label{fig:us-mobility_f}
\end{subfigure}
  \begin{subfigure}{.32\linewidth}
  \centering
  \frame{\includegraphics[width=\linewidth]{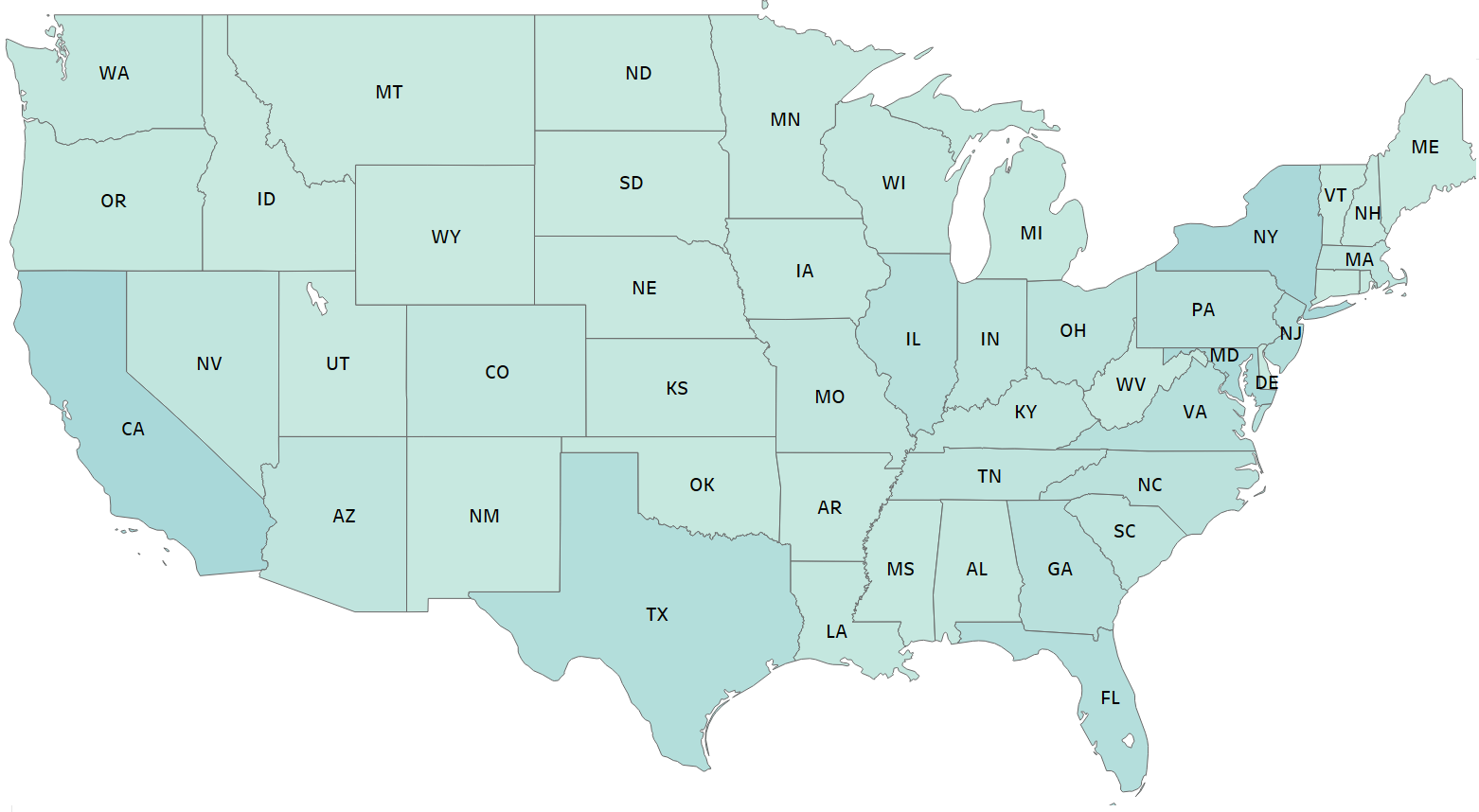}}
  \caption{May}
  \label{fig:sfig4}
\end{subfigure}
\caption{Inter-state user mobility in the U.S. for Jan-May, 2020.}
\label{fig:us-mobility}
\end{figure*}

\noindent\textbf{Inter-Region Mobility.}
%-----------------
\texttt{Mega-COV} can be exploited to generate \textit{responsive maps} where end users can check mobility patterns between different regions over time. In particular, geolocation information can show mobility patterns between regions. As an illustration of this use case, albeit in a static form, we provide Figure~\ref{fig:us-mobility_a} where we show how users move between U.S. states. We can also exploit \texttt{Mega-COV} to show inter-state mobility during a given window of time.\footnote{Here, due to increased posting in 2020, we normalize the number of visits between states by the total number of all tweets posted during a given month.} Figures~\ref{fig:us-mobility_b}-~\ref{fig:us-mobility_f} present user mobility between U.S. states. The figure shows a clear change from higher mobility in January and February to much less activity in March, April, and May. Clear differences can be seen in key states where the pandemic has hit hard such as New York (NY), California (CA), and Washington State (WA). We provide visualizations of mobility patterns for a number of countries where the pandemic has hit (sometimes hard), as follows: Brazil, Canada, Italy, Saudi Arabia, and the United Kingdom.

\noindent\textbf{Intra-Region Mobility.} We also use information in \texttt{Mega-COV} to map each user to a single home region (i.e., city, state/province, and country). We follow Geolocation literature~\cite{roller2012supervised,graham2014world,han2016twitter,do2018twitter} in setting a condition that a user must have posted at least 10 tweets from a given region. However, we also condition that at least $60\%$ of all user tweets must have been posted from the same region. We use the resulting set of users whose home location we can verify to map user weekly mobility \textit{within} their own city, state, and country exclusively for both Canada and the U.S. as illustrating examples. We provide the related visualization in supplementary material under ``\textbf{User Weekly \textit{Intra}-Region Mobility}".

% ------------- New York -----------

\begin{figure*}[]
\center
\begin{subfigure}{.35\linewidth}
\centering
\frame{\includegraphics[width=\linewidth]{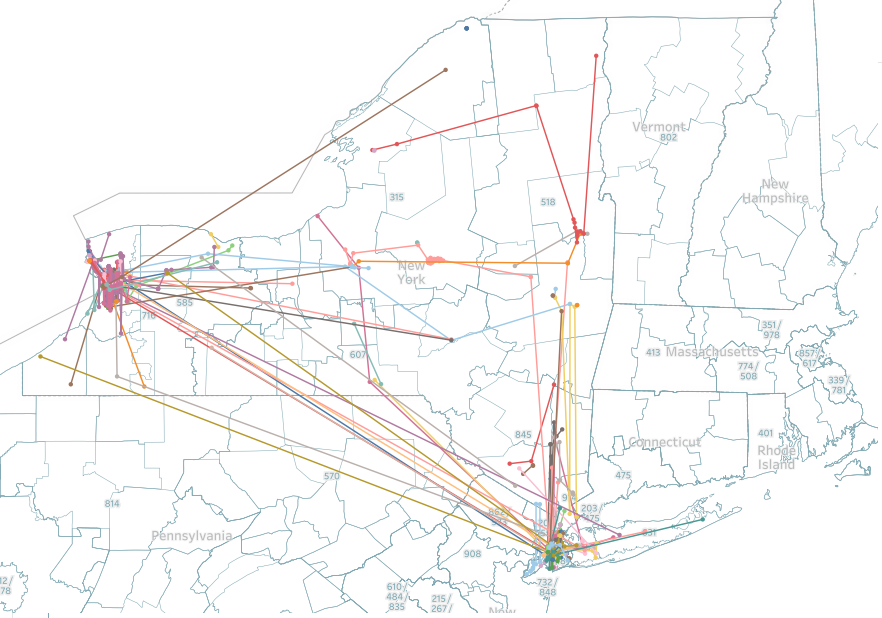}}
  \caption{January (3,949 users)}
  \label{fig:sfig1}
\end{subfigure}
\begin{subfigure}{.34\linewidth}
\centering
\frame{\includegraphics[width=\linewidth]{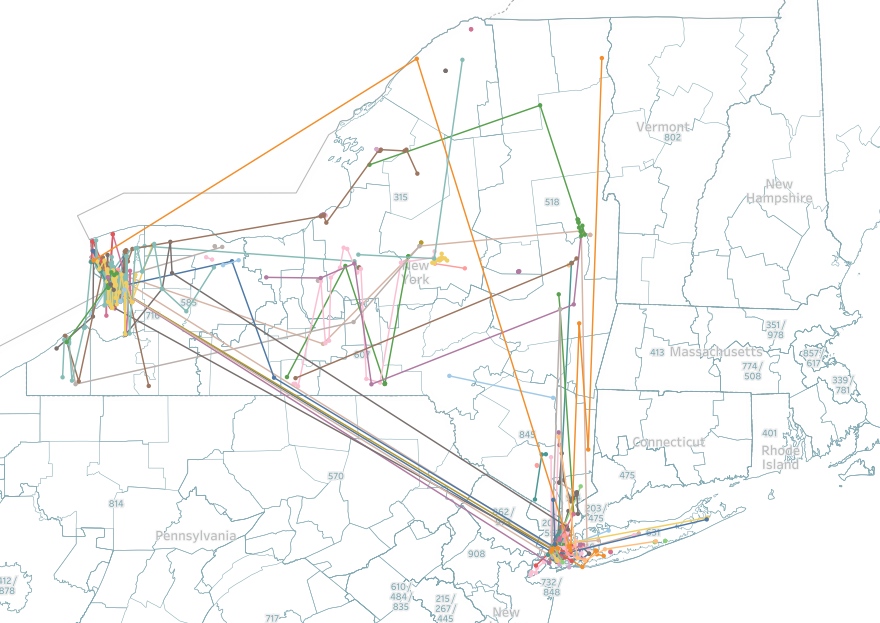}}
  \caption{February (4,500 users)}
  \label{fig:sfig1}
\end{subfigure}\\
\begin{subfigure}{.33\linewidth}
\centering
\frame{\includegraphics[width=\linewidth]{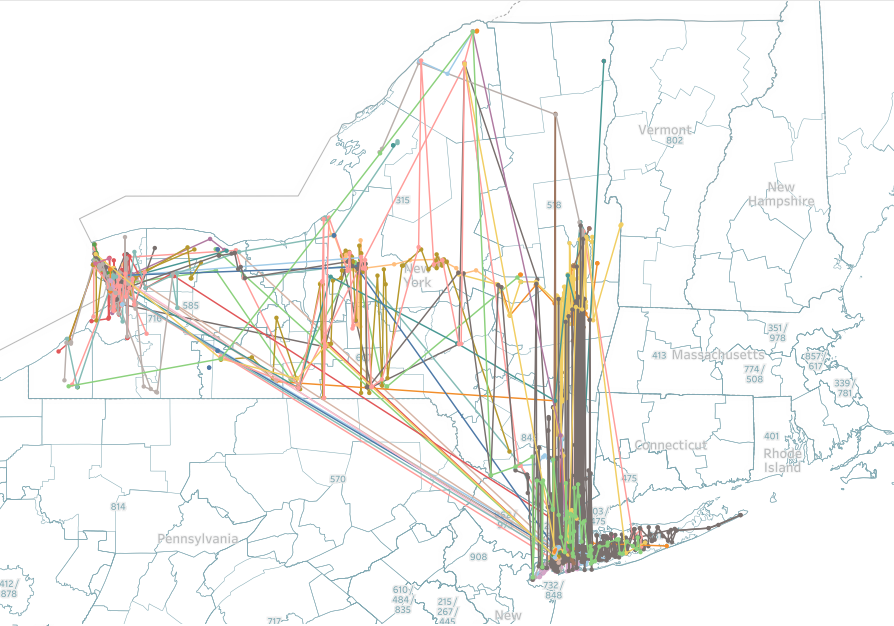}}
  \caption{March (5,145 users)}
  \label{fig:sfig1}
\end{subfigure}
\begin{subfigure}{.34\linewidth}
\centering
\frame{\includegraphics[width=\linewidth]{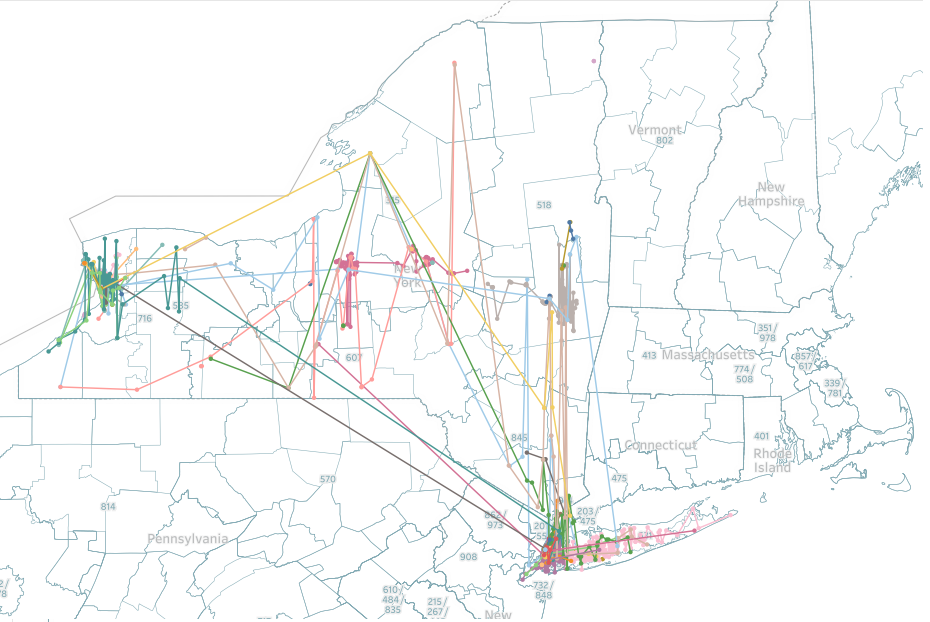}}
  \caption{April (1,870 users)}
  \label{fig:sfig1}
\end{subfigure}
\caption{User monthly mobility within New York State.}
\label{fig:mob_ny}
\end{figure*}

%---------Barazil
\begin{figure*}[]
\center
\begin{subfigure}{.33\linewidth}
  \centering
  \frame{\includegraphics[width=\linewidth]{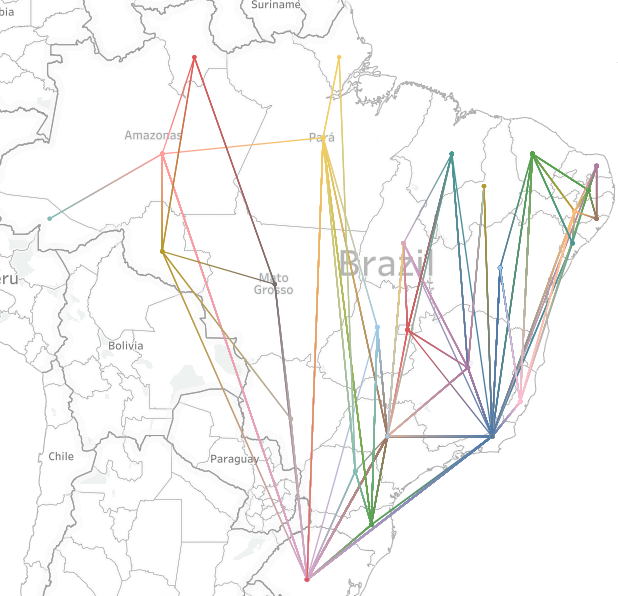}}
  \caption{Inter-States}
  \label{fig:sfig2}
\end{subfigure}
\begin{subfigure}{.32\linewidth}
  \centering
  \frame{\includegraphics[width=\linewidth]{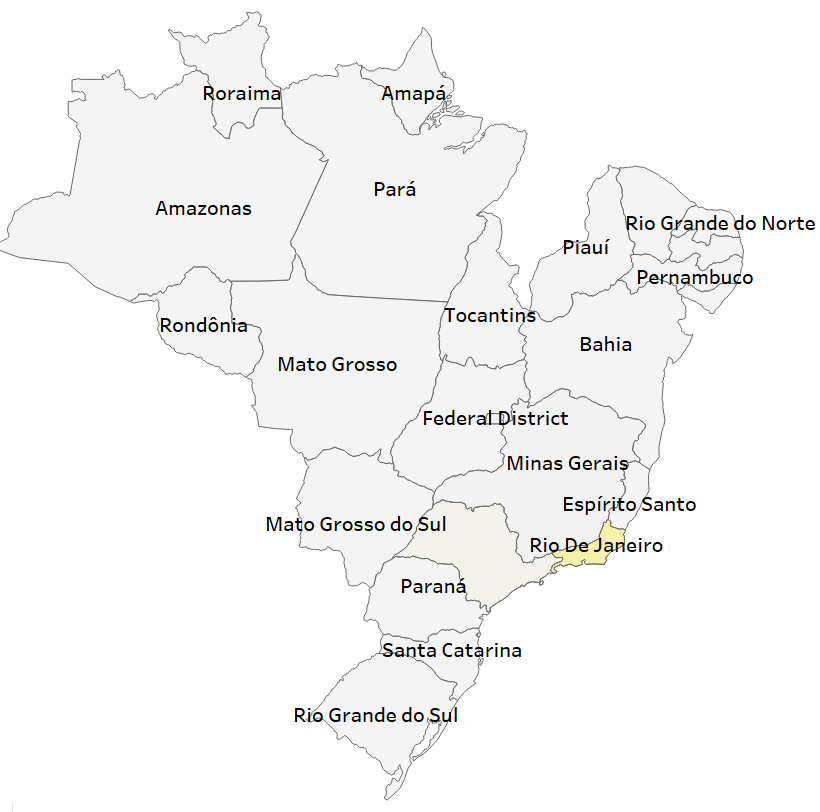}}
  \caption{January}
  \label{fig:sfig3}
\end{subfigure}
\begin{subfigure}{.32\linewidth}
  \centering
  \frame{\includegraphics[width=\linewidth]{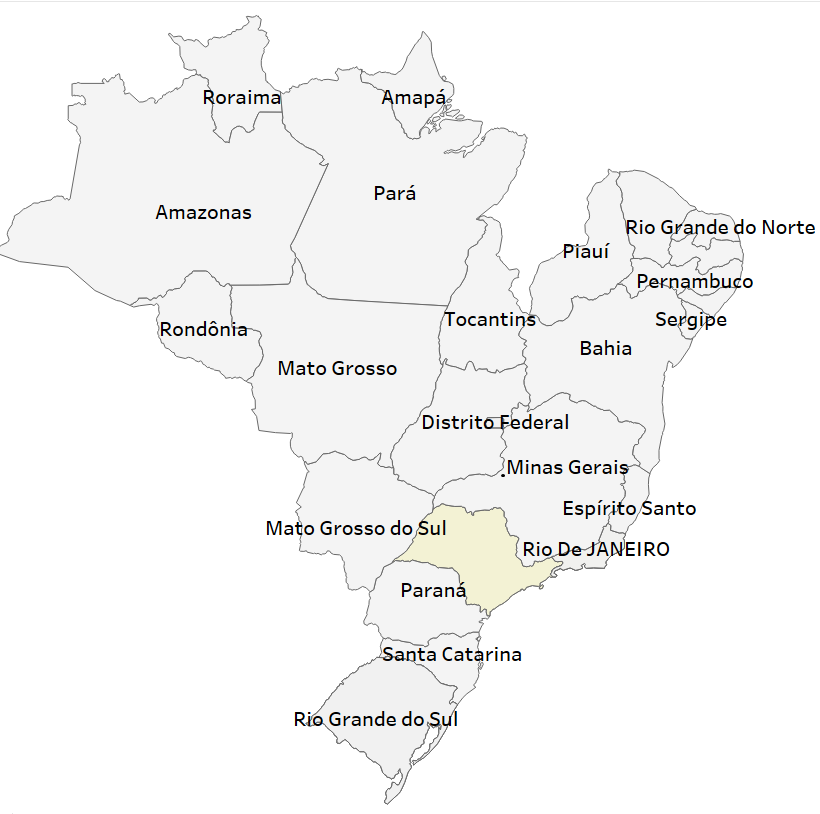}}
  \caption{February}
  \label{fig:sfig4}
\end{subfigure}\\
\begin{subfigure}{.32\linewidth}
  \centering
  \frame{\includegraphics[width=\linewidth]{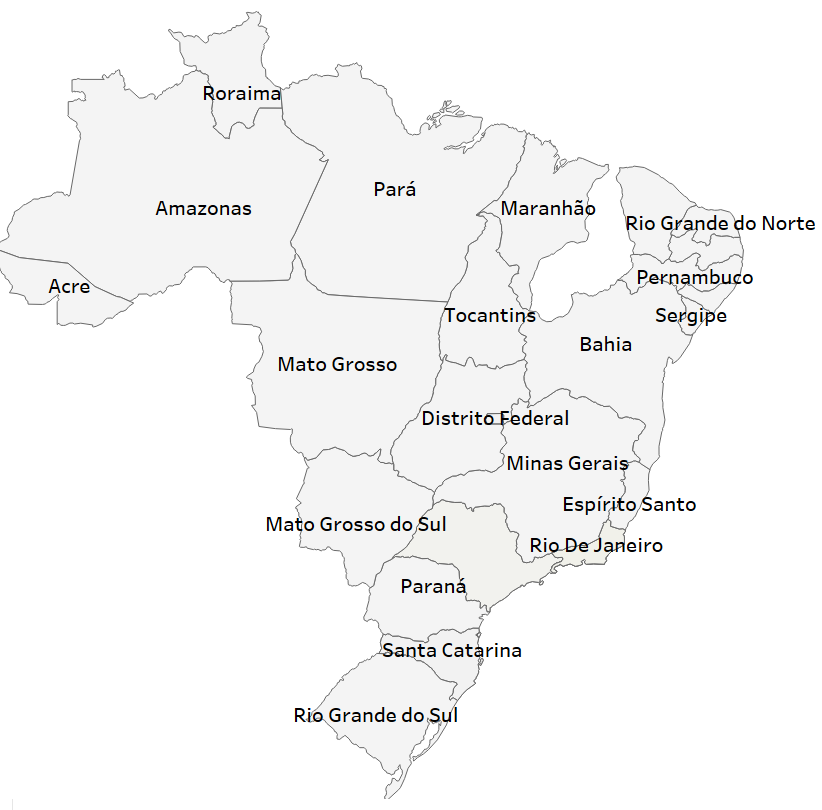}}
  \caption{March}
  \label{fig:sfig4}
\end{subfigure}
\begin{subfigure}{.32\linewidth}
  \centering
  \frame{\includegraphics[width=\linewidth]{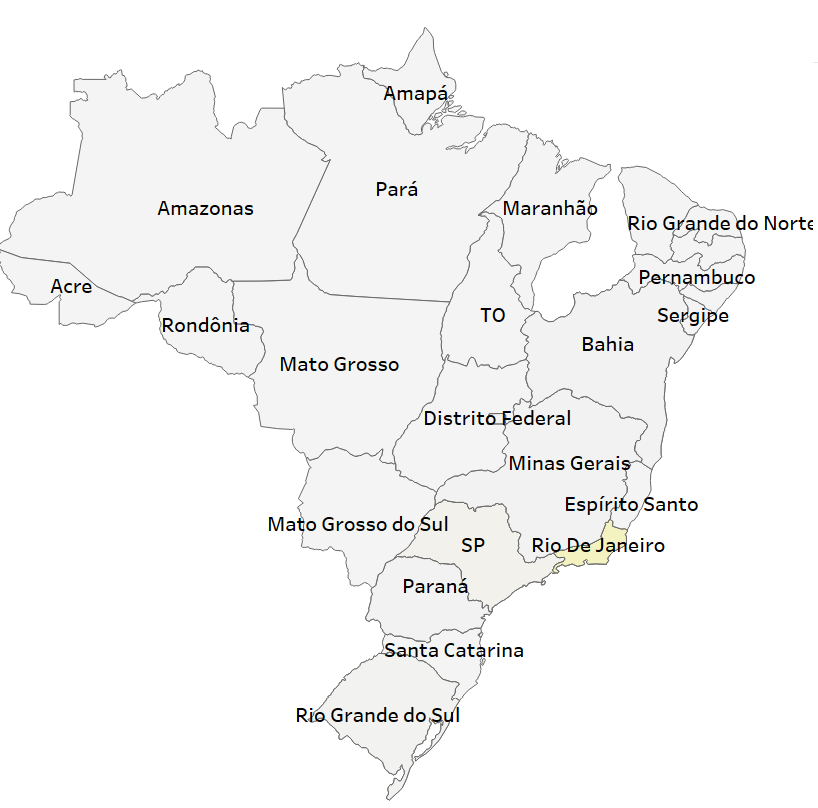}}
  \caption{April}
  \label{fig:sfig4}
\end{subfigure}
  \begin{subfigure}{.32\linewidth}
  \centering
  \frame{\includegraphics[width=\linewidth]{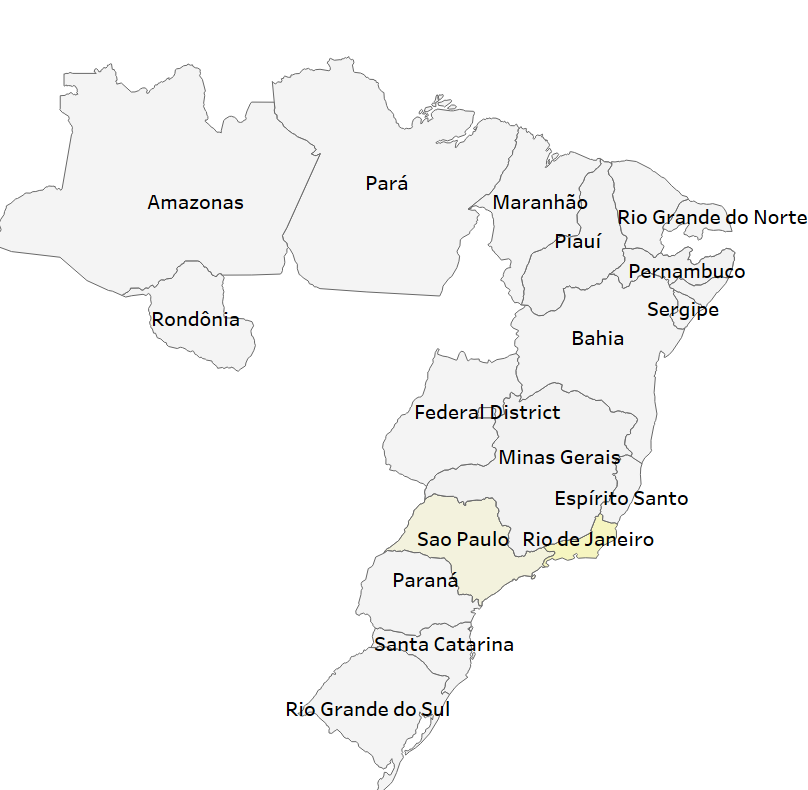}}
  \caption{May}
  \label{fig:sfig4}
\end{subfigure}
\caption{User mobility between Brazil states (estados) during Jan-May 2020.}
\label{fig:brazil_mob}
\end{figure*}
%---------Canada
\begin{figure*}[]
\center
\begin{subfigure}{.31\linewidth}
  \centering
  \frame{\includegraphics[width=\linewidth]{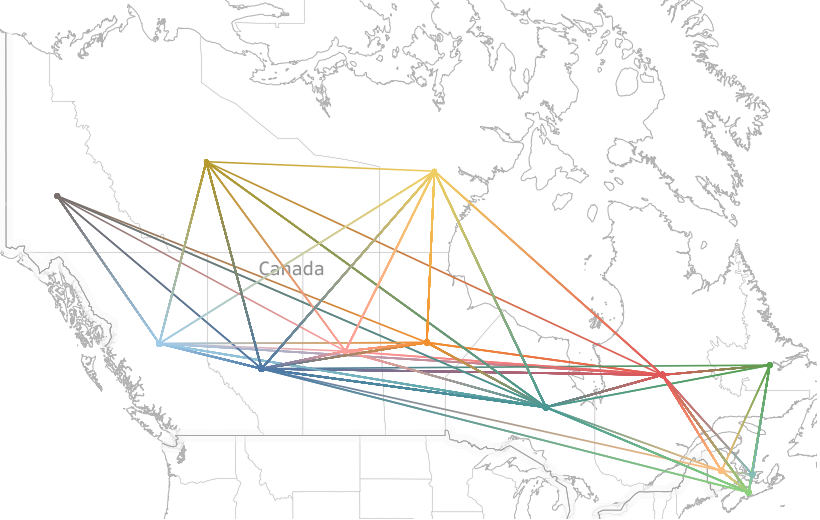}}
  \caption{Inter-provinces}
  \label{fig:sfig1}
\end{subfigure}
% \begin{subfigure}{.25\linewidth}
%   \centering
%   \frame{\includegraphics[width=\linewidth]{image/maps/UK_states_movement_2020_b12.png}}
%   \caption{Inter-States mobility}
%   \label{fig:sfig2}
% \end{subfigure}
\begin{subfigure}{.32\linewidth}
  \centering
  \frame{\includegraphics[width=\linewidth]{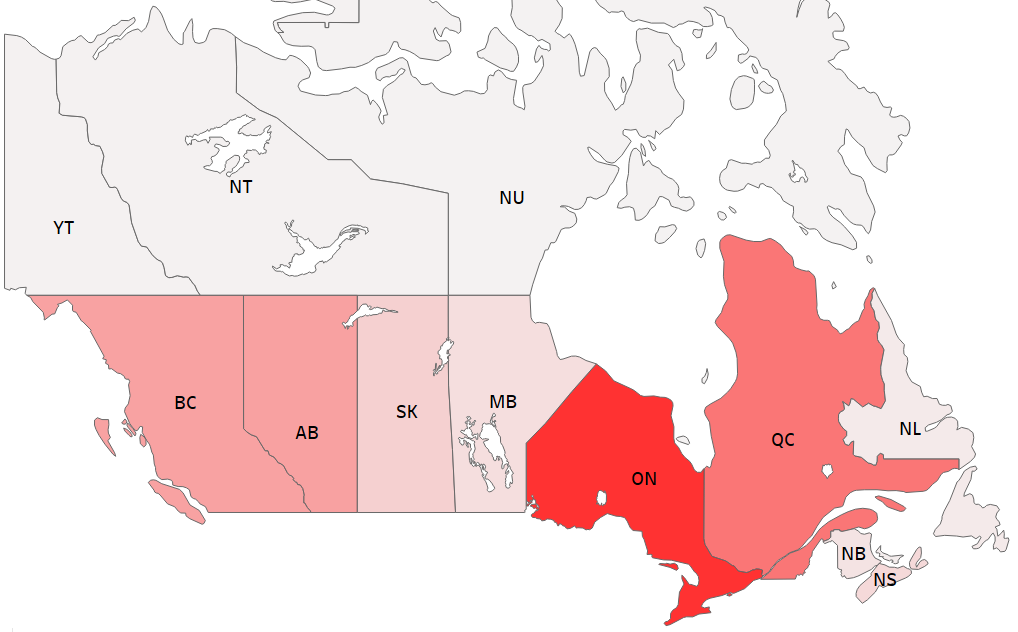}}
  \caption{January}
  \label{fig:sfig3}
\end{subfigure}
\begin{subfigure}{.32\linewidth}
  \centering
  \frame{\includegraphics[width=\linewidth]{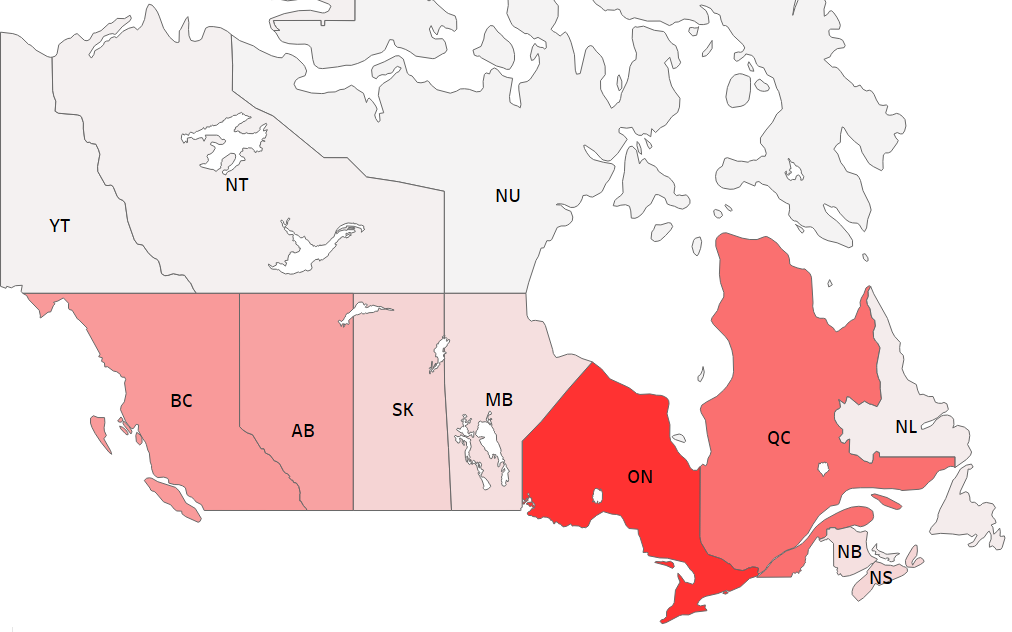}}
  \caption{February}
  \label{fig:sfig4}
\end{subfigure}\\
\begin{subfigure}{.32\linewidth}
  \centering
  \frame{\includegraphics[width=\linewidth]{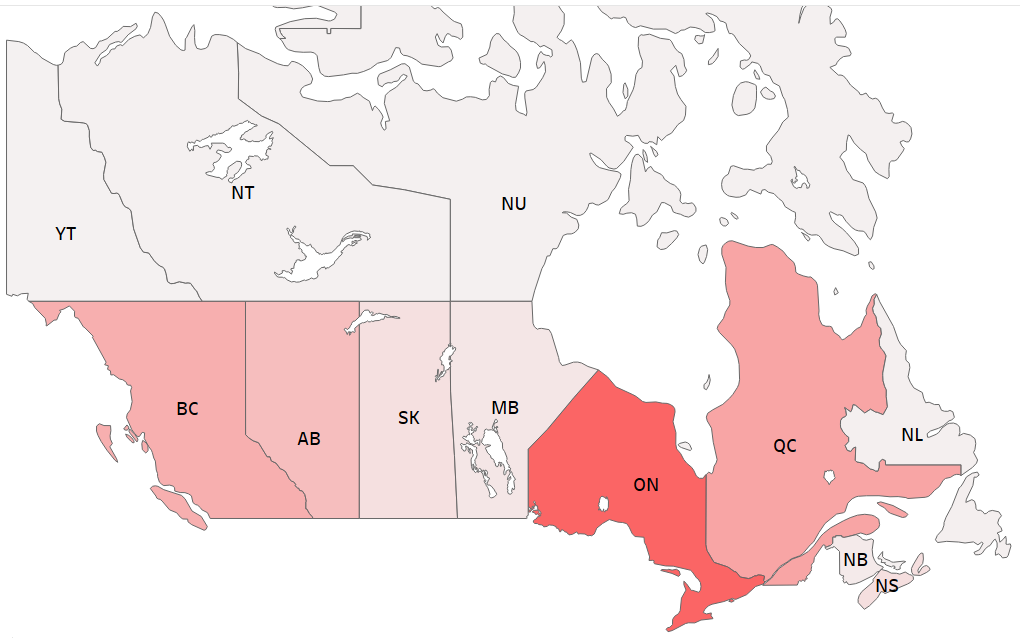}}
  \caption{March}
  \label{fig:sfig4}
\end{subfigure}
\begin{subfigure}{.32\linewidth}
  \centering
  \frame{\includegraphics[width=\linewidth]{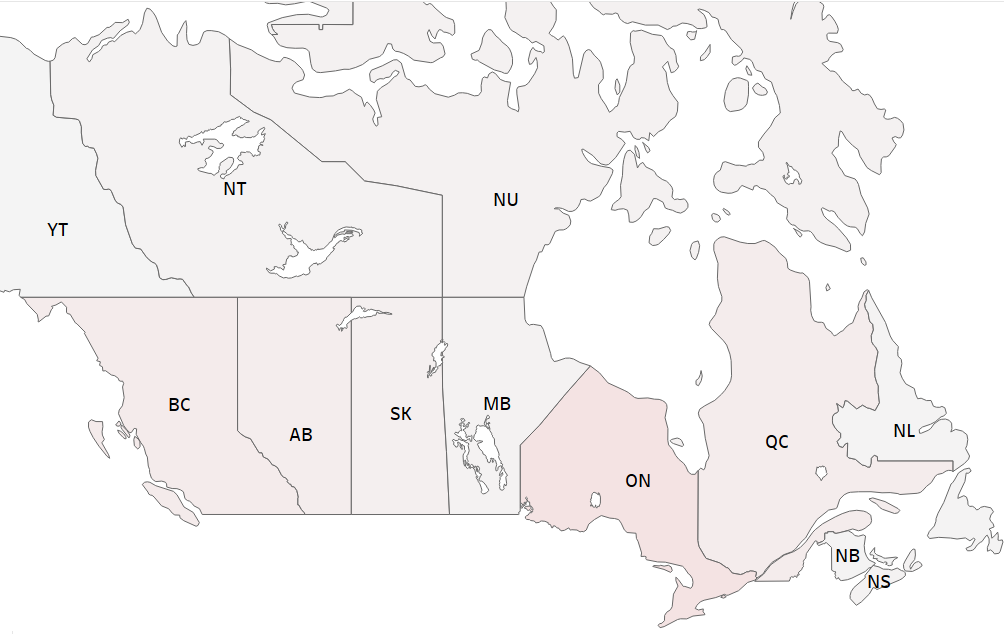}}
  \caption{April}
  \label{fig:sfig4}
\end{subfigure}
  \begin{subfigure}{.32\linewidth}
  \centering
  \frame{\includegraphics[width=\linewidth]{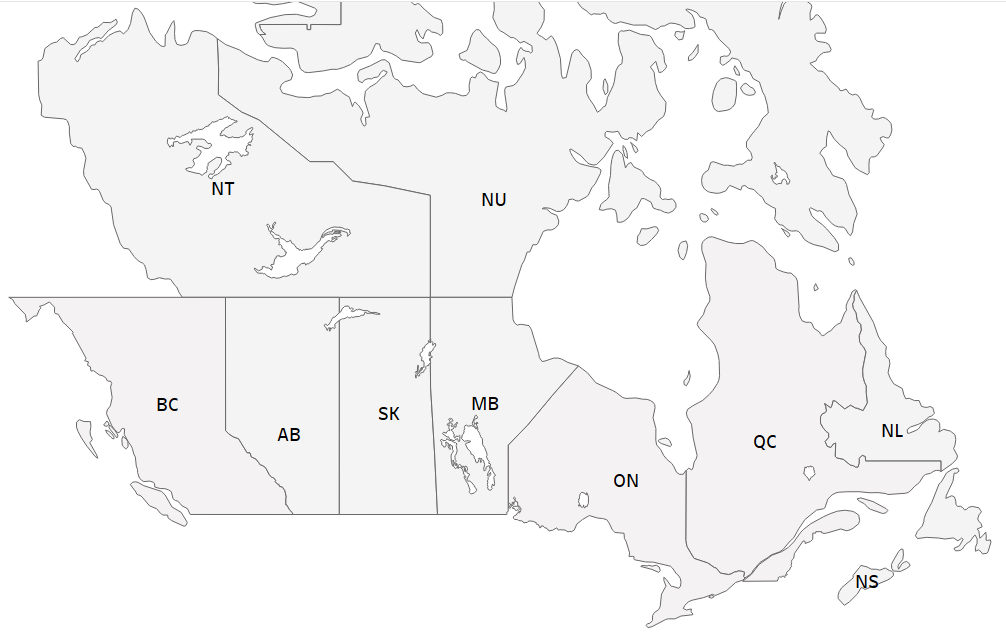}}
  \caption{May}
  \label{fig:sfig4}
\end{subfigure}
\caption{User mobility between Canada Provinces during Jan-May 2020}
\label{fig:ca_mob}
\end{figure*}
%---------Italy
\begin{figure*}[]
\center
% \begin{subfigure}{.25\linewidth}
%   \centering
%   \frame{\includegraphics[width=\linewidth]{image/maps/Italy_cities_movement_2020_b12.png}}
%   \caption{January 2020}
%   \label{fig:sfig1}
% \end{subfigure}\\
\begin{subfigure}{.295\linewidth}
  \centering
  \frame{\includegraphics[width=\linewidth]{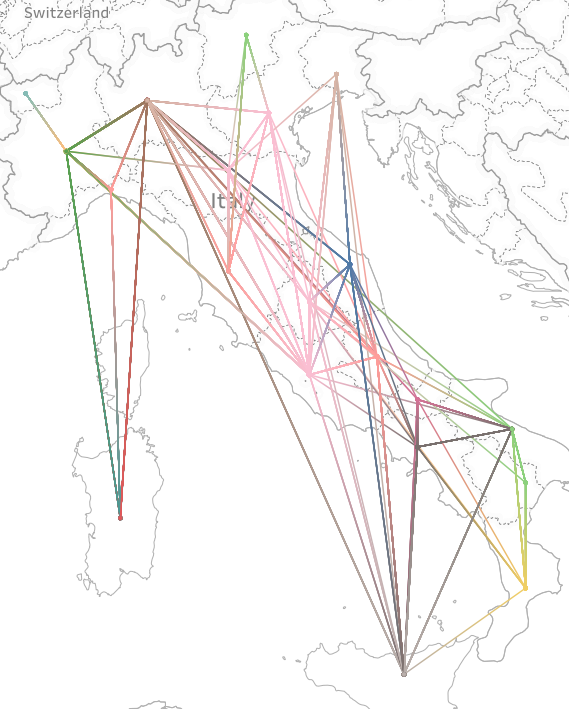}}
  \caption{Inter-Regions}
  \label{fig:sfig2}
\end{subfigure}
\begin{subfigure}{.32\linewidth}
  \centering
  \frame{\includegraphics[width=\linewidth]{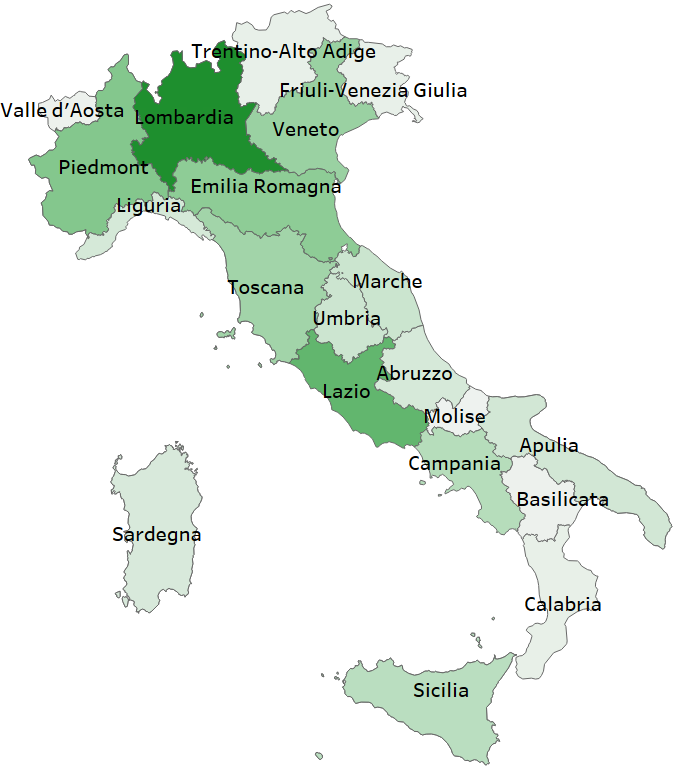}}
  \caption{January}
  \label{fig:sfig3}
\end{subfigure}
\begin{subfigure}{.32\linewidth}
  \centering
  \frame{\includegraphics[width=\linewidth]{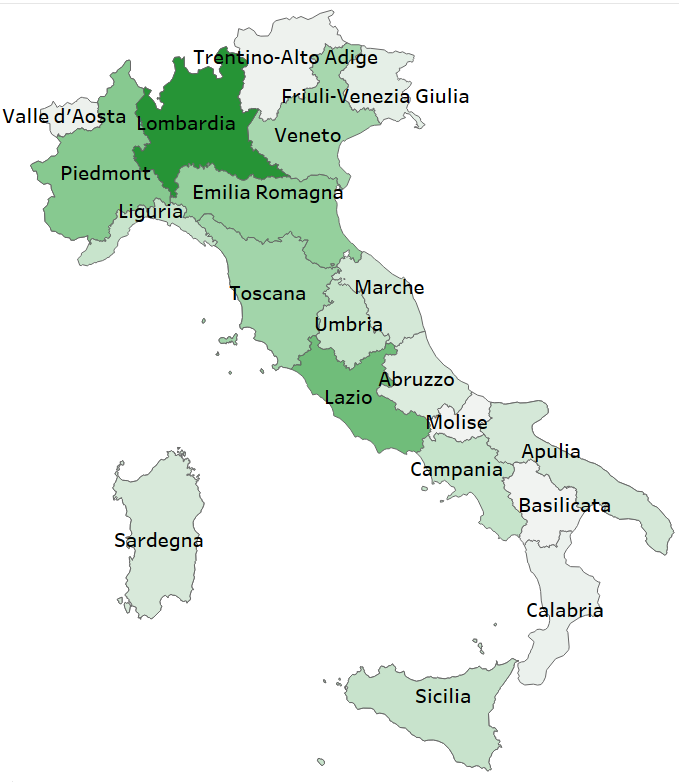}}
  \caption{February}
  \label{fig:sfig4}
\end{subfigure}
\begin{subfigure}{.32\linewidth}
  \centering
  \frame{\includegraphics[width=\linewidth]{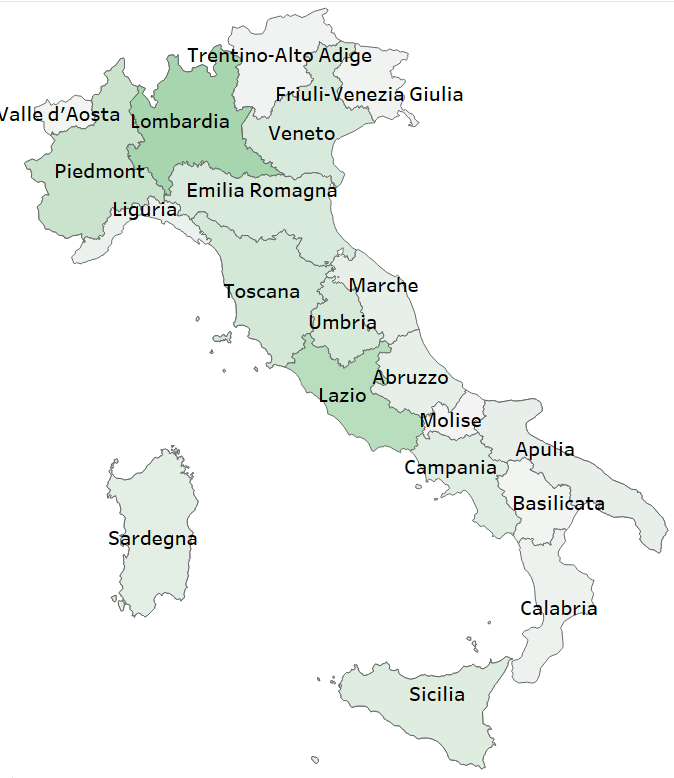}}
  \caption{March}
  \label{fig:sfig4}
\end{subfigure}
\begin{subfigure}{.32\linewidth}
  \centering
  \frame{\includegraphics[width=\linewidth]{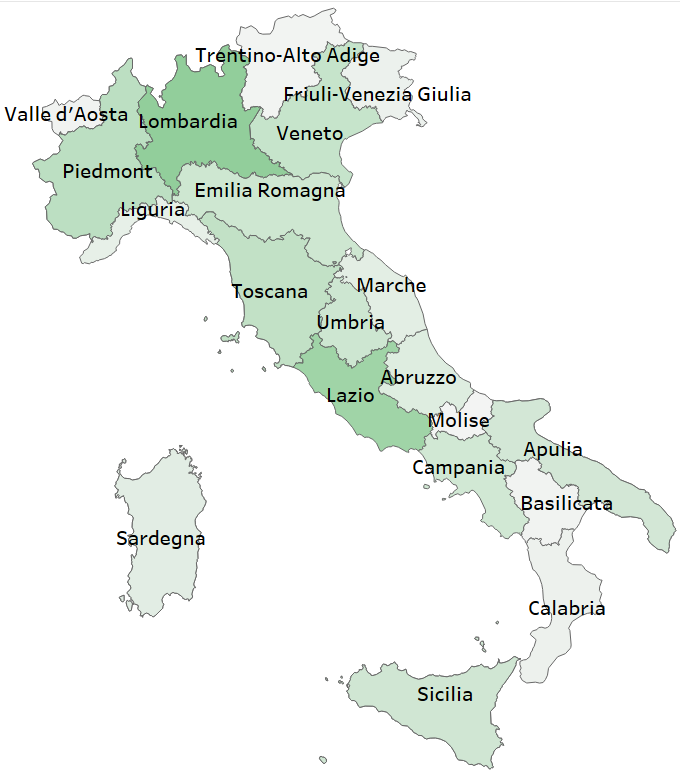}}
  \caption{April}
  \label{fig:sfig4}
\end{subfigure}
  \begin{subfigure}{.32\linewidth}
  \centering
  \frame{\includegraphics[width=\linewidth]{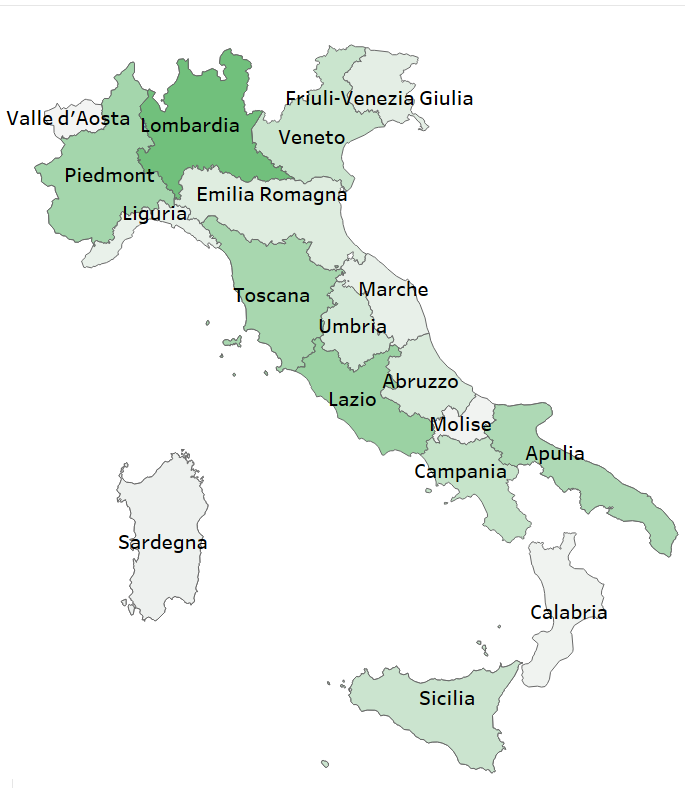}}
  \caption{May}
  \label{fig:sfig4}
\end{subfigure}
\caption{User mobility between Italy regions (regioni) during Jan-May 2020.}
\label{fig:italy_mob}
\end{figure*}
%---------ksa
\begin{figure*}[]
\center
\begin{subfigure}{.33\linewidth}
  \centering
  \frame{\includegraphics[width=\linewidth]{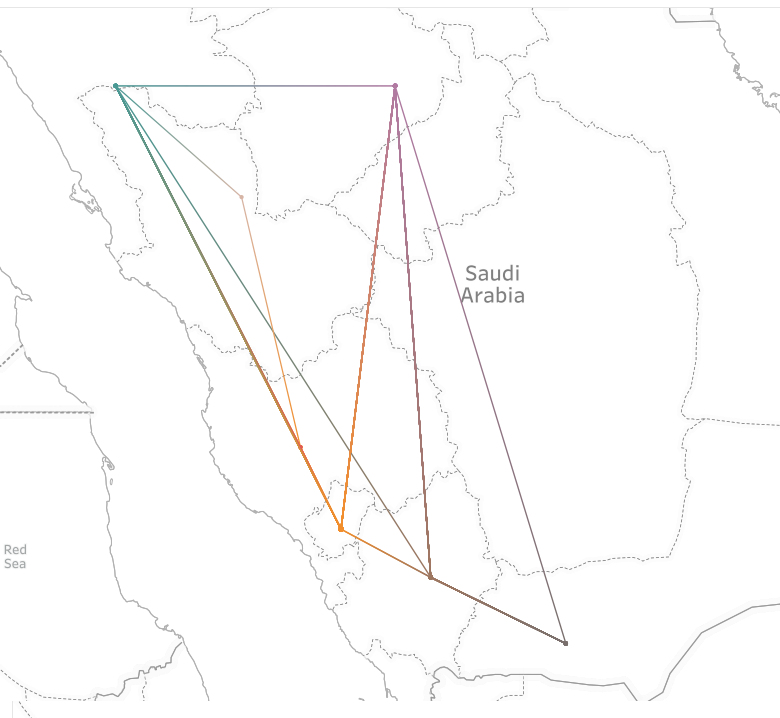}}
  \caption{Inter-Regions mobility}
  \label{fig:sfig2}
\end{subfigure}
\begin{subfigure}{.32\linewidth}
  \centering
  \frame{\includegraphics[width=\linewidth]{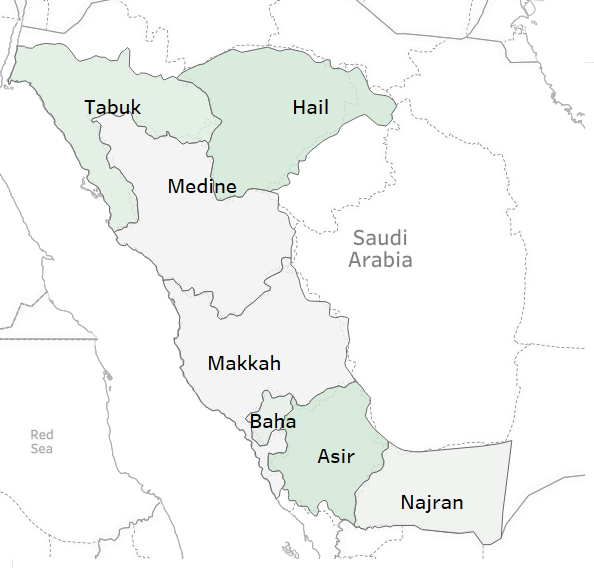}}
  \caption{January}
  \label{fig:sfig3}
\end{subfigure}
\begin{subfigure}{.32\linewidth}
  \centering
  \frame{\includegraphics[width=\linewidth]{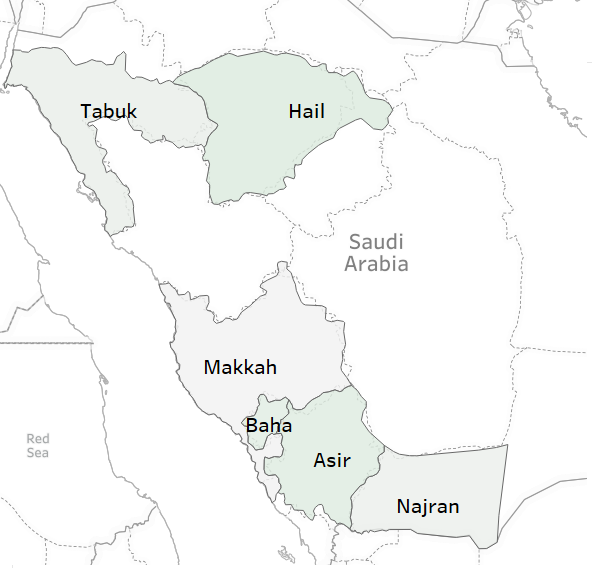}}
  \caption{February}
  \label{fig:sfig4}
\end{subfigure}\\
\begin{subfigure}{.32\linewidth}
  \centering
  \frame{\includegraphics[width=\linewidth]{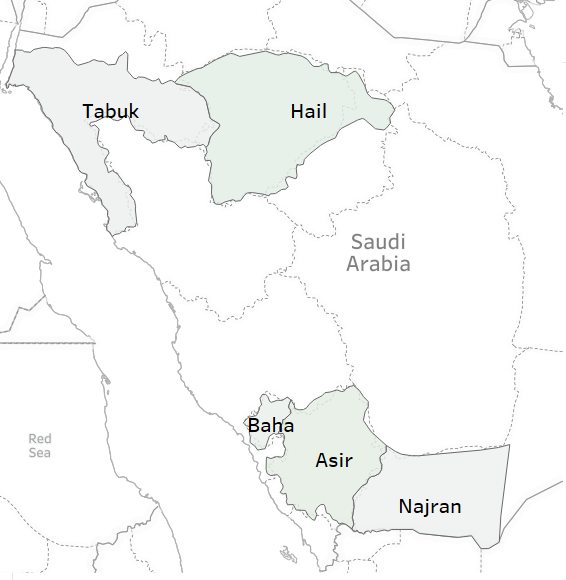}}
  \caption{March}
  \label{fig:sfig4}
\end{subfigure}
\begin{subfigure}{.32\linewidth}
  \centering
  \frame{\includegraphics[width=\linewidth]{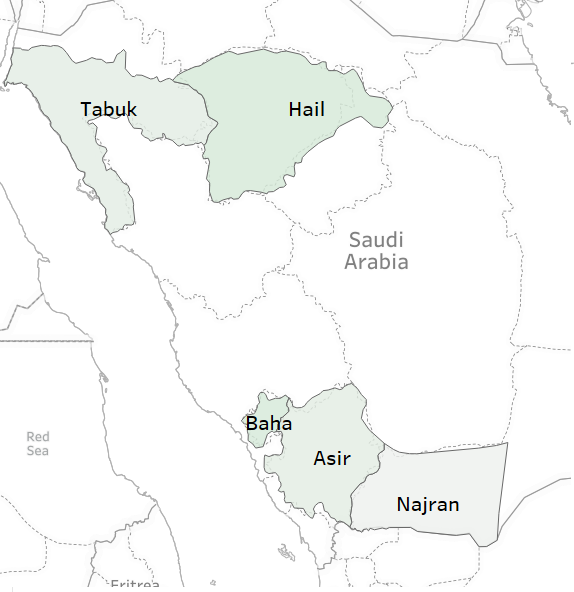}}
  \caption{April}
  \label{fig:sfig4}
\end{subfigure}
  \begin{subfigure}{.32\linewidth}
  \centering
  \frame{\includegraphics[width=\linewidth]{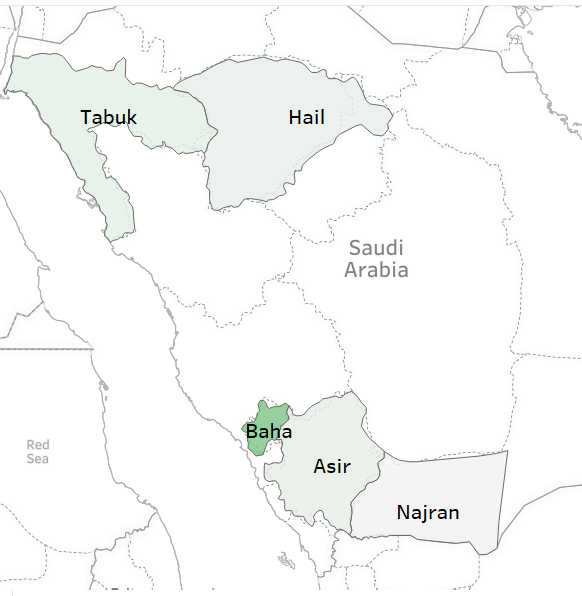}}
  \caption{May}
  \label{fig:sfig4}
\end{subfigure}
\caption{User mobility between Saudi Arabia regions during Jan-May 2020.}
\label{fig:ksa_mob}
\end{figure*}
%------------
%---------UK
\begin{figure*}[]
\center
\begin{subfigure}{.21\linewidth}
  \centering
  \frame{\includegraphics[width=\linewidth]{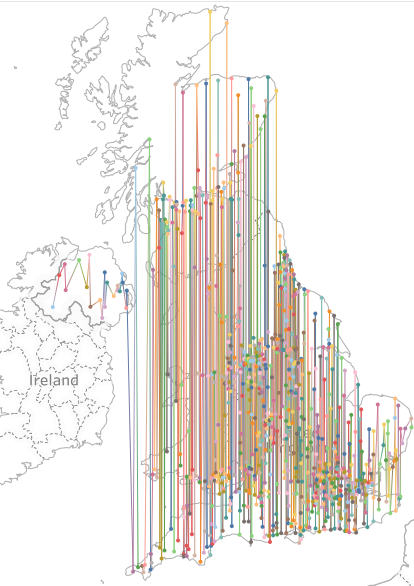}}
  \caption{Inter-cities}
  \label{fig:sfig1}
\end{subfigure}
\begin{subfigure}{.21\linewidth}
  \centering
  \frame{\includegraphics[width=\linewidth]{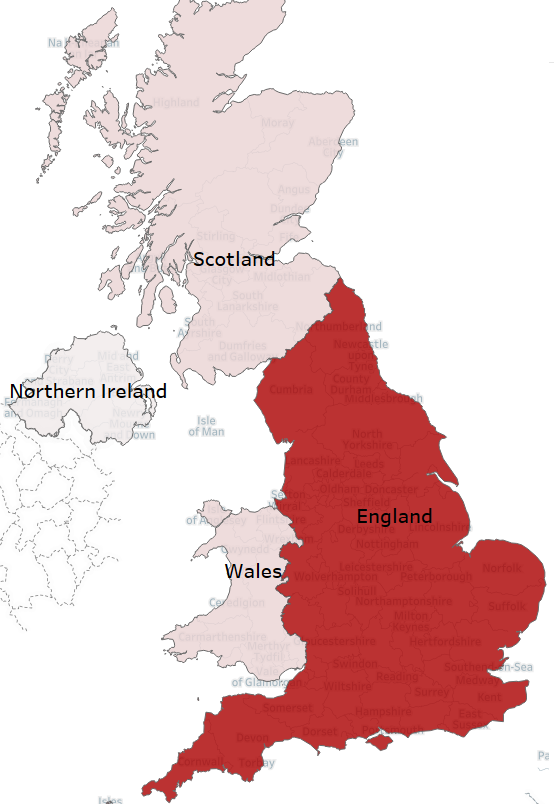}}
  \caption{January}
  \label{fig:sfig3}
\end{subfigure}
\begin{subfigure}{.21\linewidth}
  \centering
  \frame{\includegraphics[width=\linewidth]{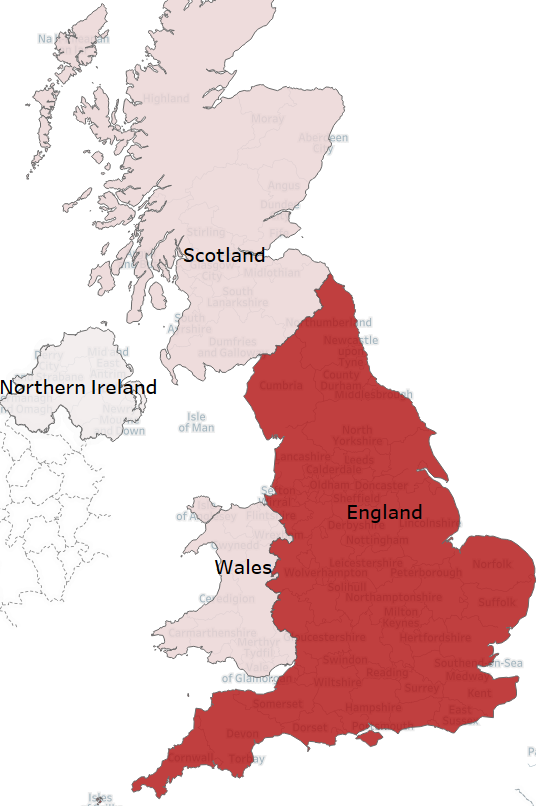}}
  \caption{February}
  \label{fig:sfig4}
\end{subfigure}\\
\begin{subfigure}{.21\linewidth}
  \centering
  \frame{\includegraphics[width=\linewidth]{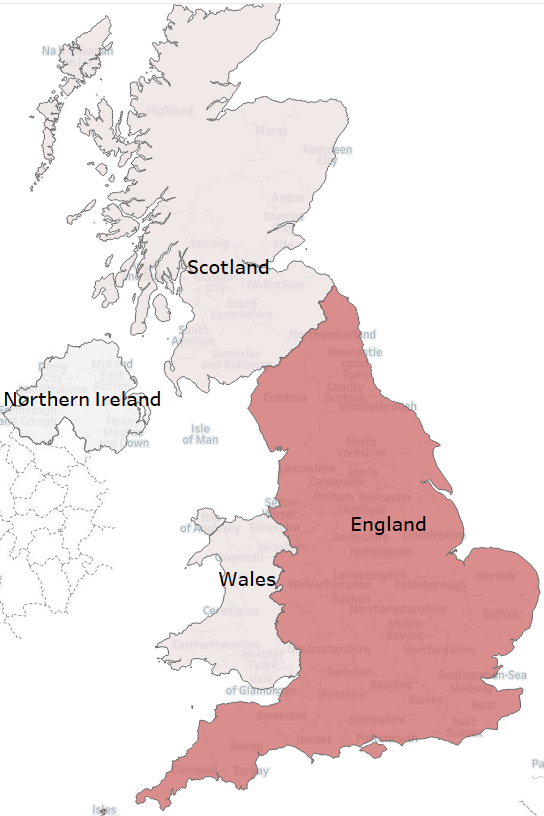}}
  \caption{March}
  \label{fig:sfig4}
\end{subfigure}
\begin{subfigure}{.21\linewidth}
  \centering
  \frame{\includegraphics[width=\linewidth]{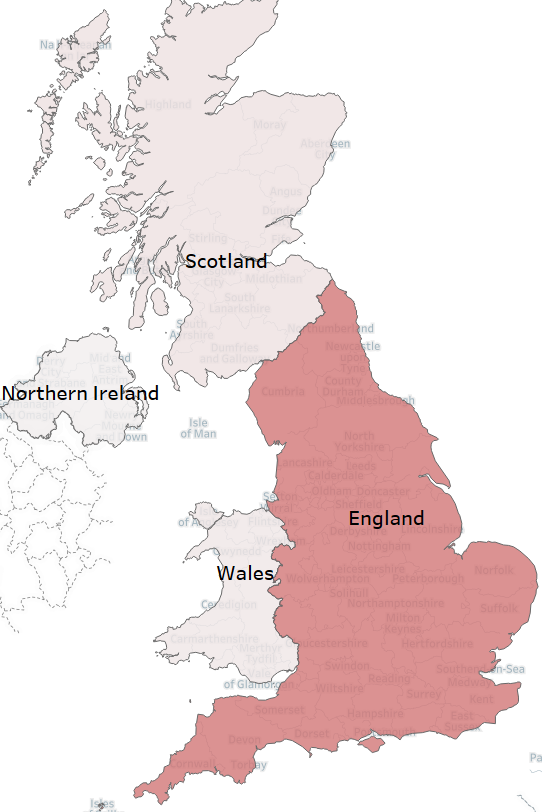}}
  \caption{April}
  \label{fig:sfig4}
\end{subfigure}
  \begin{subfigure}{.21\linewidth}
  \centering
  \frame{\includegraphics[width=\linewidth]{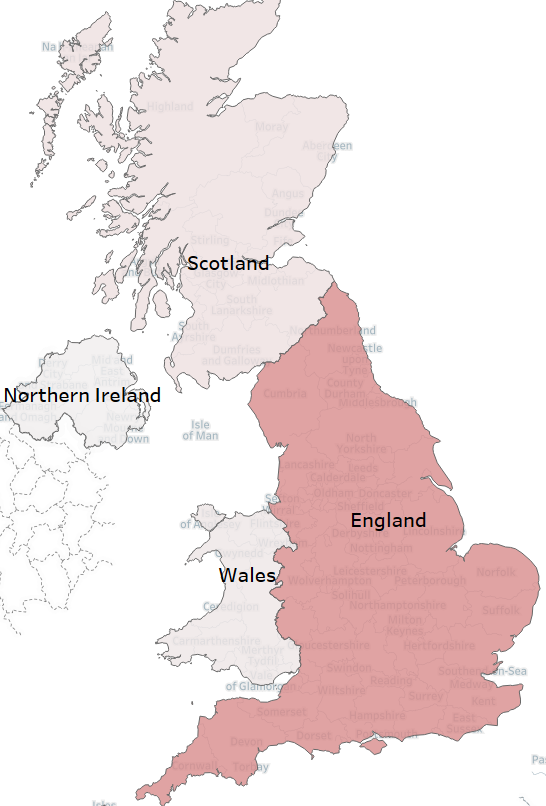}}
  \caption{May}
  \label{fig:sfig4}
\end{subfigure}
\caption{User mobility between United Kingdom  counties during Jan-May 2020.}
\label{fig:uk_mob}
\end{figure*}

\begin{table*}[]
\begin{adjustbox}{width=\columnwidth*2}
\begin{tabular}{ll|ll|ll}
\hline
\multicolumn{2}{c|}{\textbf{Replies}}    & \multicolumn{2}{c|}{\textbf{RT}}          & \multicolumn{2}{c}{\textbf{Tweets}}      \\ \hline
\textbf{Lang}                 & \textbf{Frequency} & \textbf{Lang}                 & \textbf{Frequency} & \textbf{Lang}                 & \textbf{Frequency} \\ \hline
Hebrew (he)           & 337,237   & Hebrew (he)           & 322,351   & Hebrew (he)           & 504,327   \\
Croatian (hr)         & 242,772   & Croatian (hr)         & 198,764   & Croatian (hr)         & 243,601   \\
Maltese (mt)          & 94,695    & Maltese (mt)          & 85,054    & Maltese (mt)          & 145,395   \\
Dzongkha (dz)         & 64,063    & Slovak(sk)            & 77,846    & Slovak(sk)            & 131,544   \\
Bahasa Indonesia (id) & 46,463    & Latin (la)            & 66,061    & Latin (la)            & 104,488   \\
Bosnian (bs)          & 43,066    & Bahasa Indonesia (id) & 61,295    & Bahasa Indonesia (id) & 100,561   \\
Slovak(sk)            & 36,662    & Bosnian (bs)          & 45,276    & Bosnian (bs)          & 54,200    \\
Swahili (sw)          & 21,803    & Swahili (sw)          & 28,122    & Dzongkha (dz)         & 46,950    \\
Azerbaijani (az)      & 20,242    & Dzongkha (dz)         & 26,853    & Swahili (sw)          & 42,076    \\
Latin (la)            & 13,030    & Quechua (qu)          & 22,559    & Malay (ms)            & 32,967    \\
Albanian (sq)         & 12,878    & Malay (ms)            & 19,511    & Quechua (qu)          & 31,175    \\
Xhosa (xh)            & 11,936    & Esperanto (eo)        & 19,397    & Albanian (sq)         & 30,361    \\
Irish (ga)            & 8,607     & Kinyarwanda (rw)      & 19,371    & Kinyarwanda (rw)      & 30,080    \\
Malagasy (mg)         & 7,727     & Azerbaijani (az)      & 19,182    & Azerbaijani (az)      & 29,507    \\
Quechua (qu)          & 7,449     & Javanese (jv)         & 18,180    & Javanese (jv)         & 29,121    \\
Kinyarwanda (rw)      & 7,427     & Albanian (sq)         & 17,904    & Esperanto (eo)        & 29,019    \\
Esperanto (eo)        & 6,755     & Xhosa (xh)            & 14,886    & Kurdish (ku)          & 24,259    \\
Malay (ms)            & 6,683     & Irish (ga)            & 14,807    & Afrikaans (af)        & 22,871    \\
Assamese (as)         & 6,442     & Kurdish (ku)          & 14,475    & Volapük (vo)          & 21,840    \\
Volapük (vo)          & 6,245     & Galician (gl)         & 13,337    & Irish (ga)            & 21,151   \\ \hline   
\end{tabular}
\end{adjustbox}
\caption{Top 20 languages detected by langid in \texttt{Mega-COV V0.1} which were not detected by twitter, broken by tweets, retweets, and replies.}
\label{tab:top-langid-cats}
\end{table*}

\begin{figure*}[]
\center
\begin{subfigure}{\linewidth}
\centering
\frame{\includegraphics[width=\linewidth]{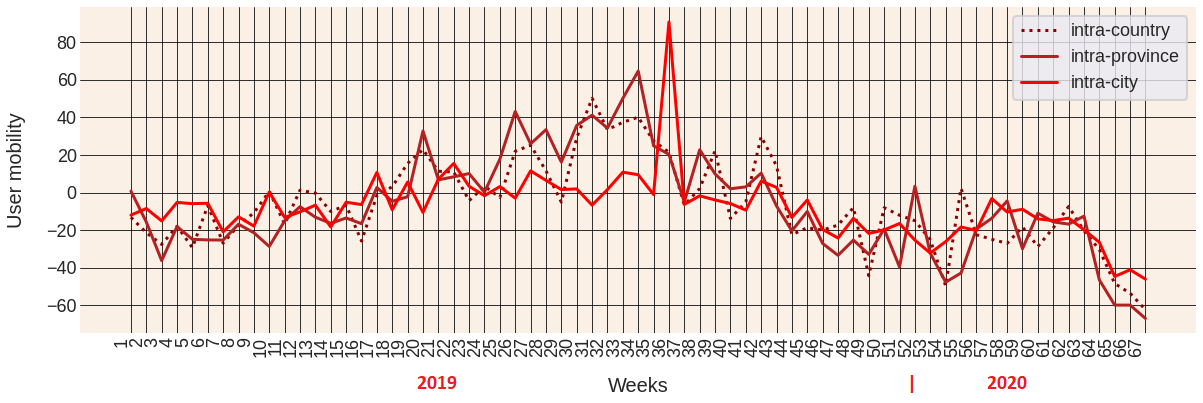}}
  \caption{Canada users}
  \label{fig:sfig1}
\end{subfigure}\\
\begin{subfigure}{\linewidth}
\centering
  \frame{\includegraphics[width=\linewidth]{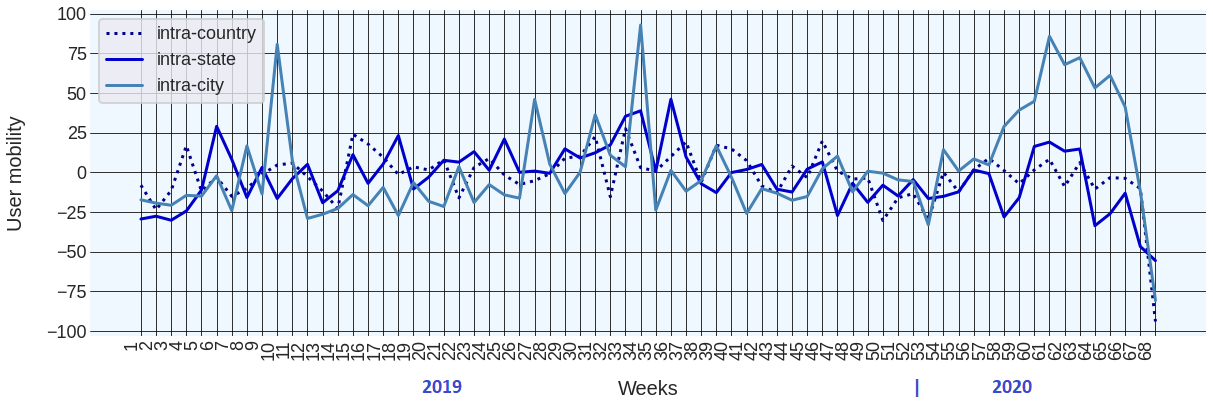}}
  \caption{U.S. users}
  \label{fig:sfig3}
\end{subfigure}
\caption{Canadian and American user \textit{weekly} mobility during 2019-2020. Each point (a week) is modeled as a mobility distance from weekly average mobility in 2019.}
\label{fig:weekly_mob}
\end{figure*}

%%%%%%%%%%%%%%%%%%
\newpage
%---------
\subsection{\textbf{User Weekly Intra-Region Mobility}}\label{sec:intra_regon}
%--------------------------------
We can also visualize user mobility as a distance from an average mobility score on a weekly basis. Namely, we calculate an \textit{average weekly mobility score} for the year 2019 using geo-tag information (longitude and latitude) and use it as a \textit{baseline} against which we plot user mobility for each week of 2019 and 2020 up until April. In general, we observe a drop in user mobility in Canada starting from mid-March. For U.S. users, we notice a very high mobility surge starting around end of February and early March, only waning down the last week of March and continuing in April as shown in Figure~\ref{fig:weekly_mob}. For both the U.S. and Canada, we hypothesize the surge in early March (much more noticeable in the U.S.) is a result of people moving back to their hometowns, returning from travels, moving for basic needs stocking, etc.

%%%%%%%%%%%%%%%%%%

\section{\textbf{ COVID-Relevance Model }}\label{sec:append:models}
%---------

\subsection{Dataset}
We randomly sample 200K tweets from the English data in~\newcite{chen2020covid} and a maximum of 100K from each of the rest of languages. For languages where there is $< 100$K tweets, we take all data. For the negative class, we extract data from Jan-Nov, 2019 from \texttt{Mega-COV}. For each language, we take roughly the same number of tweets we sampled for the positive class. Table~\ref{tab:append:relevance_data} shows the distribution of the positive class data from~\newcite{chen2020covid}.

% \usepackage{graphicx}
\begin{table}[]
\begin{adjustbox}{width=8cm}

\renewcommand{\arraystretch}{1.3}
{\footnotesize
\begin{tabular}{cccccccc}
\toprule
\textbf{lang} & \textbf{\#tweets} & \textbf{lang} & \textbf{\#tweets} & \textbf{lang} & \textbf{\#tweets} & \textbf{lang} & \textbf{\#tweets} \\ \toprule
en            & 200K              & ar            & 76K               & uk            & 6.6K              & sr            & 838               \\ 
es            & 100K              & ru            & 50K               & no            & 5.5K              & bg            & 739               \\ 
th            & 100K              & lt            & 44.7K             & eu            & 4.8K              & dv            & 634               \\ 
fr            & 100K              & pl            & 40.6K             & cy            & 4.3K              & pa            & 450               \\ 
in            & 100K              & fa            & 32.6K             & ne            & 4K                & my            & 277               \\ 
ja            & 100K              & ro            & 32.5K             & lv            & 3.2K              & ps            & 244               \\ 
pt            & 100K              & sv            & 24.2K             & mr            & 3K                & am            & 229               \\ 
it            & 100K              & fi            & 24K               & iw            & 2.6K              & ckb           & 190               \\ 
und           & 100K              & vi            & 22.7K             & ml            & 2.4K              & sd            & 144               \\ 
tr            & 100K              & et            & 21.3K             & hu            & 2.2K              & km            & 128               \\ 
tl            & 100K              & ur            & 20.3K             & te            & 2.1K              & lo            & 47                \\ 
de            & 100K              & ht            & 16.2K             & gu            & 1.8K              & hy            & 34                \\ 
zh            & 100K              & da            & 16.2K             & bn            & 1.5K              & ka            & 23                \\ 
ca            & 100K              & sl            & 13.5K             & kn            & 1.4K              & bo            & 15                \\ 
nl            & 100K              & cs            & 13.3K             & or            & 1.2K              & ug            & 5                 \\ 
ko            & 100K              & ta            & 13.1K             & is            & 1.2K              &               &                   \\ 
hi            & 97.4K             & el            & 10.7K             & si            & 1.2K              &               &                   \\ 
\toprule    
\end{tabular}}
\end{adjustbox}
\caption{\small {Distribution of language in the COVID-Relevance training data for the positive (i.e., \textit{related}) classe.}   }\label{tab:append:relevance_data}
\end{table}
\newpage
\section{\textbf{COVID-Misinformation Detection  }}\label{sec:append:misinfo}
%%%%%%%%%
%%%%%%%%%%%%%%%%%%%%%%%%%%%%
\citet{cui2020coaid} present a \textbf{C}ovid-19 he\textbf{A}thcare m\textbf{I}sinformation \textbf{D}ataset
(CoAID), with diverse COVID-19 healthcare misinformation, including fake news on websites and social platforms, along with related user engagement (i.e., tweets and replies) about such news. CoAID includes $3,235$ news articles and claims, $294,692$ user engagement, and $851$ social platform posts about COVID-19. The topics of  CoAID include: \textit{\{COVID-19, coronavirus, pneumonia, flu9, lock down, stay home, quarantine and ventilator\}}. The dataset is collected from December 1, 2019 to July 1, 2020 and is organized as follows:  

\begin{itemize}
    \item \textbf{News Articles.} To collect the \textit{true} news (not fake), $9$ reliable media outlets were identified. These include World Health Organization\footnote{\url{https://www.who.int/}} and the U.S. National Institute of Health\footnote{\url{https://www.nih.gov/}}, for example. To collect \textit{fake} news, $6$ fact-checking websites were used (e.g. LeadStories\footnote{\url{https://leadstories.com/hoax-alert/}}, PolitiFact\footnote{\url{https://www.politifact.com/coronavirus/}}).
    \item \textbf{Claims.} The true and fake claims (i.e., news with one or two sentences) were collected using: (1) the official WHO website,\footnote{\url{https://www.who.int/}} (2) WHO official Twitter account,\footnote{\url{https://twitter.com/who}} and (3) the medical news today website\footnote{\url{https://www.medicalnewstoday.com}}. 
    \item \textbf{User Engagement.} Queries based on the true and fake articles and claims were used to build a dataset of user engagement from Twitter where the goal was to acquire the tweets discussing the news in question and related Twitter replies.  \\
\end{itemize}

\begin{table}[]
\begin{adjustbox}{width=8cm}

\renewcommand{\arraystretch}{1.3}
{\footnotesize
\begin{tabular}{lcccccc}
\toprule
\multicolumn{1}{c}{} & \multicolumn{6}{c}{\textbf{News}}                                                             \\\cline{2-7}
\multicolumn{1}{c}{} & \multicolumn{3}{c}{\textbf{Fake}}             & \multicolumn{3}{c}{\textbf{True}}             \\ \cline{2-7}
\multicolumn{1}{c}{} & \textbf{TRAIN} & \textbf{DEV} & \textbf{TEST} & \textbf{TRAIN} & \textbf{DEV} & \textbf{TEST} \\\toprule
\textbf{CoAID*}        & 669            & 84           & 84            & 2,172          & 272          & 272           \\ 
\textbf{ReCOVery}    & 532            & 66           & 66            & 1,091          & 136          & 136           \\
%\textbf{FakeCovid}   & 1,908          & 238          & 238           & 2,180          & 273          & 273           \\
\toprule
                     & \multicolumn{6}{c}{\textbf{Tweets}}                                                           \\ \cline{2-7}
                     & \multicolumn{3}{c}{\textbf{Fake}}                      & \multicolumn{3}{c}{\textbf{True}}                      \\\toprule
                     & \textbf{TRAIN} & \textbf{DEV} & \textbf{TEST} & \textbf{TRAIN} & \textbf{DEV} & \textbf{TEST} \\\toprule
\textbf{CoAID}       & 8,072          & 1,009        & 1,009         & 110,076        & 13,759       & 13,759        \\
\textbf{ReCOVery}    & 18,272         & 2,284        & 2,284         & 86,437         & 10,805       & 10,805        \\
%\textbf{FakeCovid}   & -              & -            & -             & -              & -            & -         \\
\toprule    
\end{tabular}}
\end{adjustbox}
\caption{\small {Statistics of CoAID, ReCOVery, and FakeCovid datasets across the data splits. For \textbf CoAID$^*$, we merge the claim and news.}   }\label{tab:data_splits}
\end{table}

%%%%%%%%%%%%%%%%%%%%%%%%%%%%%%%%%%%%%%%%%%%%%%%%%%%%%%%%%%%%%%%%%%%%  
\begin{table}[H]
 	
\small
\begin{adjustbox}{width=8cm}
\renewcommand{\arraystretch}{1.7}{
\small
\centering
\begin{tabular}{llcccc}

\toprule
 \multirow{2}{*}{\textbf{Data}}  & \multirow{2}{*}{\textbf{ Model}} & \multicolumn{2}{c}{\textbf{DEV}}     & \multicolumn{2}{c}{\textbf{TEST}}       \\  \cline{3-6} 
       &      &   \textbf{Acc.} & \textbf{ F1}         & \textbf{Acc.}    & \textbf{F1}        \\\toprule
  %\multirow{2}{*}{\rotatebox[origin=c]{0}{\textbf{\colorbox{blue!10}{Baseline}}}}&  

 % &    &      AraBERT        & $73.07$&			$67.10$&			$72.59$&			$67.05$ \\ \cline{2-7}

%%%%%%%%%%%%%%%%%%%%%%%%%%%%%%%%%%%%%%%%%%%%%%%%%%%%%%%%%%%%%%%%%  

%\multirow{6}{*}{\rotatebox[origin=c]{0}{\textbf{\colorbox{blue!10}{News}}}}   &
%    &  Baseline   &  $76.40$ & $86.62$   &  $76.40$ & $86.62$  \\  \cdashline{2-6} 

%LSTM   &$88.42$&$92.51$&$90.89$&$92.94$\\ \cdashline{2-6} 
 \multirow{3}{*}{\small \textbf{CoAID}}       & mBERT           & $98.88$&$	98.45$& {$\bf 97.47$} &$\textbf{96.48}$\\

 &     XLM-R\textsubscript{Base}  &   $98.31$& 97.64 &$96.35$&$94.74$\\
     &      XLM-R\textsubscript{Large}        & {$\bf 99.16$}&{$\bf 98.84$}& 96.91 &\ {$95.66$}\\
  \toprule

%%%%%%%%%%%%%%%%%%%%%%%%%%%%%%%%%%%%%%%%%%%%%%%%%%%%%%%%%%%%%%%%%  

%\multirow{4}{*}{(2b)}
   % &  Baseline   &  $67.15$ & $80.35$   &   $67.15$ & $80.35$ \\  \cdashline{2-6} 
  
  % LSTM   &$71.98$&$82.22$&$73.84$&$82.49$\\   \cdashline{2-6} 
  \multirow{3}{*}{ \bf{ReCOVery} }      & mBERT           &$86.76$&$84.14$&$85.64$&$82.73$\\ 

 &     XLM-R\textsubscript{Base}  &   $85.78$&$83.56$&$87.13$&$85.01$\\ 
     &      XLM-R\textsubscript{Large}        &   {$\bf 88.73$}& {$\bf 86.36$}& {$ \bf 88.12$}& {$\bf 85.91$}\\

\cline{2-6}
%%%%%%%%%%%%%%%%%%%%%%%%%%%%%%%%%%%%%%%%%%%%%%%%%%%%%%  
  
 \toprule 

%%%%%%%%%%%%%%%%%%%%%%%%%%%%%%%%%%%%%%%%%%%%%%%%%%%%%%%%%%%%%%%%%  

%\multirow{3}{*}{\rotatebox[origin=c]{0}{\ {\colorbox{blue!10}{News}}}}   &
  % &      Baseline   &  $73.03$ & $84.41$   &   $73.03$ & $84.41$  \\  \cdashline{2-6} 
 
 %  LSTM    & $80.96$&$88.28$&$78.94$&$85.77$\\  \cdashline{2-6} 
 \multirow{3}{*}{\small \textbf{CoAID+ReCOV.}}   & mBERT     & {$\bf 93.39$} & {$\bf91.41$}&$92.11$&$89.66$\\
    &     XLM-R\textsubscript{Base}  &$92.50$&$89.90$&$91.04$&$87.79$\\
  &      XLM-R\textsubscript{Large}           &  $93.21$& $90.86$& {$\bf 92.83$}&{$ \bf 90.37$}\\
  
\cline{2-6}

  \toprule

\end{tabular}
} \end{adjustbox}

\caption{\small Results of our fake news detector models on the DEV and TEST splits of CoAID  and ReCOVery news articles datasets.}%, and  \textbf{(d)} $5$x AraNews$^+$.
\label{tab:append:res-fake-news}
\end{table}

\end{document}